\documentclass[preprint,aps,pre,letterpaper,superscriptaddress,longbibliography]{revtex4-1}
\usepackage{amsmath,amssymb,bm}
\usepackage{mathrsfs}
\usepackage{textcomp}
\usepackage{commath}
\usepackage{mathtools}

\usepackage{hyperref}
\hypersetup{
    colorlinks=true,
    linkcolor=blue,
}
\usepackage{placeins}
\usepackage[final]{graphicx} 
\usepackage{rotating}
\usepackage{epstopdf}
\usepackage{float}
\usepackage[lofdepth=1,lotdepth]{subfig}
\usepackage[normalem]{ulem}

\usepackage[usenames]{color} 

\newcommand{\MDG}[1]{\textcolor{blue}{}} 
\newcommand{\MDGrevise}[1]{\textcolor{black}{#1}} 
\newcommand{\XZ}[1]{\textcolor{red}{}} 
\newcommand{\XZrevise}[1]{\textcolor{black}{#1}}
\renewcommand{\sout}[1]{}

\newcommand{\Ca}{\mathrm{Ca}}

\graphicspath{{figures/}}  

\makeatletter

\newcommand{\numtoRoman}[1]{\expandafter\@slowromancap\romannumeral #1@}
\makeatother

\begin{document}

\title{Multiplicity of stable orbits for deformable prolate capsules in shear flow}
\author{Xiao Zhang}
\affiliation{
Department of Chemical and Biological Engineering\\
University of Wisconsin-Madison, Madison, WI 53706-1691
}
\author{Michael D. Graham}\email{Corresponding author. E-mail: mdgraham@wisc.edu}
\affiliation{
Department of Chemical and Biological Engineering\\
University of Wisconsin-Madison, Madison, WI 53706-1691
}
\date{\today}

\begin{abstract}
This work investigates the orbital dynamics of a fluid-filled deformable prolate capsule in unbounded simple shear flow at zero Reynolds number using direct simulations. The motion of the capsule is simulated using a model that incorporates shear elasticity, area dilatation, and bending resistance. Here the deformability of the capsule is characterized by the nondimensional capillary number $\Ca$, which represents the ratio of viscous stresses to elastic restoring stresses on the capsule. For a capsule with small bending stiffness, at a given $\Ca$, the orientation converges over time towards a unique stable orbit independent of the initial orientation. With increasing $\Ca$, four dynamical modes are found for the stable orbit, namely, rolling, wobbling, oscillating-swinging, and swinging. On the other hand, for a capsule with large bending stiffness, multiplicity in the orbit dynamics is observed. When the viscosity ratio $\lambda \lesssim 1$, the long-axis of the capsule always tends towards a stable orbit in the flow-gradient plane, either tumbling or swinging, depending on $\Ca$. When $\lambda \gtrsim 1$, the stable orbit of the capsule is a tumbling motion at low $\Ca$, irrespective of the initial orientation. Upon increasing $\Ca$, there is a symmetry-breaking bifurcation away from the tumbling orbit, and the capsule is observed to adopt multiple stable orbital modes including nonsymmetric precessing and rolling, depending on the initial orientation. As $\Ca$ further increases, the nonsymmetric stable orbit loses existence at a saddle-node bifurcation, and rolling becomes the only attractor at high $\Ca$, whereas the rolling state coexists with the nonsymmetric state at intermediate values of $\Ca$. A symmetry-breaking bifurcation away from the rolling orbit is also found upon decreasing $\Ca$. The regime with multiple attractors becomes broader as the aspect ratio of the capsule increases, while narrowing as  viscosity ratio increases. We also report the particle contribution to the stress, which also displays multiplicity. 
\end{abstract}
\maketitle
\newpage
\section{INTRODUCTION} \label{sec:introduction}

Microcapsules, small liquid droplets enclosed by a thin solid membrane, have been of increasing significance in the bioengineering, pharmaceutics, and food industries. Examples include cell encapsulation for tumor progression monitoring \cite{Alessandri14843}, encapsulation of cosmetic active ingredients for topical application \cite{Casanova2016}, development of drug delivery systems \cite{MASUDA201123,DEGEN2014611}, and artificial food particles for aquatic filter feeders \cite{jones1974}.
 
To satisfy engineering needs, many fabrication techniques for artificial capsules have emerged. Polyelectrolyte capsules (PECs), for example, are a promising vehicle in the biomedical field to carry a variety of therapeutic molecules (peptides, proteins, etc.) for targeted delivery to a desired site in the body \cite{DeGeest_2007}. Avoidance of immune clearance (cellular uptake by macrophages), which is critical for the efficiency of carrier particles in delivery system, has been revealed to greatly depend on particle physical and chemical properties such as size, shape, deformability, and surface chemistry. A number of studies have reported that high-aspect-ratio (prolate) microparticles exhibited considerably reduced immune clearance and increased circulation half-time compared to spheres \cite{Geng2007,Champion2009}. To this end, Zan et al. \cite{Zan2015} developed a fabrication method to obtain PECs with constant surface chemistry but independently controlled size and shape by combining soft organic templates created by the particle stretching method and a modified layer-by-layer (LBL) deposition process, which may enable both a more systematic investigation on the roles of capsule properties on its efficiency and the optimization of the design of a delivery system. Furthermore, non-spherical capsules, because of a higher surface-to-volume ratio than their spherical counterparts, can be preferable to enhance the transfer of cargoes across the capsule membrane upon arrival at the target; an example is the fabrication of prolate capsules containing a calcium ion solution by Schneeweiss and Rehage \cite{Schneeweiss2005} using a microfluidic channel.  

All of the aforementioned situations involve capsule suspensions in a fluid environment, and the dynamics of these suspensions can be greatly affected by the behavior of single capsules. Therefore, it is essential to gain a comprehensive understanding of the single capsule dynamics in a flow. For example, an initially spherical deformable capsule is found to take a so-called tank-treading motion in simple shear flow with the membrane rotating periodically around its stationary elongated shape \cite{barthes-biesel_rallison_1981}. Here the deformability of the capsule is characterized by the capillary number $\Ca$, which represents the ratio of viscous stresses to elastic restoring stresses on the particle. Non-spherical capsules, such as oblate spheroids \cite{Ramanujan:1998tx,Sui2008,kessler_finken_seifert_2008,Bagchi2009,Le2010,walter-salsac-DBB-2011,Yazdani2011_2,Omori2012,Cordasco:2013hb,Wang2013,dupont_delahaye_barthes-biesel_salsac_2016} and biconcave discoids, a model for red blood cells (RBCs) \cite{Goldsmith351,Fischer894,Viallat2007PRL,Secomb2007PRL,Viallat2010PRL,FEDOSOV20102215,Fedosov:2011cf,Peng:2011gn,Yazdani2011PRE,Dupire:2012fz,Cordasco:2014go,Peng:2014jz,Sinha:2015wt}, have also been investigated extensively in both experimental and computational studies. These studies have explored a wide range of parameters for the capsule membrane mechanics and flow properties, and revealed rich and complex orbital dynamics for the capsules. Indeed, the increasing attention that oblate capsules have gained in the past decades may be attributed to the findings by several studies on RBCs \cite{Dupire:2012fz,Cordasco:2014go,Peng:2014jz,Sinha:2015wt} that an oblate spheroidal spontaneous shape has to be assumed for RBCs to maintain the stability of their biconcave shape during motions, and it provides a better prediction than other spontaneous shapes for cell dynamics in comparison with experimental observations. 

In contrast, to the best of our knowledge, there only exist a limited number of works on the motion of prolate capsules in shear flow, most of which have focused on the special case where the fluids inside and outside the capsules have the same viscosity. Walter et al. \cite{walter-salsac-DBB-2011} numerically studied the motion of an inertialess prolate capsule in shear flow with the major axis initially positioned in the shear plane. The membrane was modeled as a thin hyperelastic surface with no bending resistance. Two stable in-plane orbital motions were found: a rigid-body-like tumbling motion at low $\Ca$, in which the capsule flips continuously, and a fluid-like motion named swinging at high $\Ca$, which is similar to a tank-treading motion only with small oscillations in both the deformation and orientation of the capsule. Richer orbital dynamics have been revealed for a prolate capsule with out-of-shear-plane initial orientations. In a numerical investigation by Dupont \emph{et al.} \cite{Dupont2013}, the membrane mechanics of the capsule was described using a model that includes shear elasticity and area dilatation described by either the Skalak law \cite{SKALAK:1973tp} or the neo-Hookean law, but again no bending resistance. They showed that for any initial orientation with respect to the vorticity axis, the capsule always converges towards a unique stable long-time orbit depending on $\Ca$. At low $\Ca$, the stable orbit corresponds to a rolling motion about the vorticity axis. As $\Ca$ increases, the capsule precesses around the vorticity axis, and undergoes a so-called wobbling motion. At high $\Ca$, the capsule assumes a complex motion with oscillations about the shear plane termed oscillating-swinging, which eventually evolves into a swinging motion, as described in \cite{walter-salsac-DBB-2011}, as $\Ca$ further increases. The same qualitative motions were generally observed in the parameter regimes considered in this work, irrespective of the capsule membrane law or aspect ratio. Cordasco and Bagchi \cite{Cordasco:2013hb} later incorporated a small constant bending stiffness into the membrane for an inertialess prolate capsule with zero spontaneous curvature, and observed a transition of the stable orbit from a drift precession (rolling) to a stable precession with a tank-treading motion of the membrane (oscillating-swinging) as $\Ca$ increases, which is qualitatively similar to the observations in \cite{Dupont2013}.

The effect of a small particle inertia was considered by Wang \emph{et al.} \cite{Wang2013} for the three-dimensional orbital behavior of an isolated prolate capsule in shear flow, in which, again, bending resistance was neglected and a unity viscosity ratio between the inner and outer fluids was assumed. With increasing $\Ca$, the dominant stable orbital modes were found to be tumbling, precessing, rolling, and oscillating-swinging. In contrast to the findings in \cite{Dupont2013} and \cite{Cordasco:2013hb}, they revealed that multiple stable orbits coexist in the transition regimes, even at the same capillary number. 

None of the preceding studies regarding deformable prolate capsules have addressed the effects of bending stiffness and viscosity ratio, which may both play a nontrivial role in the orbital dynamics of the capsules. In fact, in applications, the mechanical properties of the capsule membrane can vary greatly depending on the materials and fabrication techniques, and the suspending fluid or solution and the fluid enclosed inside the capsules will generally have different viscosities. The present work aims to gain an improved understanding of the dynamics of prolate capsules in shear flow by performing a systematic numerical investigation over a  broad domain of parameter space and determining the effects of membrane deformability, bending stiffness, initial orientation, aspect ratio of the capsule and the viscosity ratio between the inner and outer fluids. In this work, we reveal parameter regimes for an inertialess deformable prolate capsule in shear flow in which either a single or multiple attractors exist, and show in particular that the orbits for a capsule lying in the shear plane and aligned with the vorticity axis are not always stable as the capsule evolves towards the equilibrium configuration. By computing the particle contribution to the stress, we illustrate how the multiplicity of stable orbits for a prolate capsule is reflected in the rheology for a suspension of such capsules in the dilute limit. The connection between orbit dynamics and rheology for capsules has been reported in a number of prior studies \cite{BARTHESBIESEL1981493,pozrikidis_1995,Ramanujan:1998tx,Pozrikidis2003,Drochon2003,Bagchi2010,Clausen2010,bagchi_kalluri_2011,takeishi_rosti_imai_wada_brandt_2019}, but only a small number have considered prolate capsules. 

The rest of the paper is organized as follows: in Section \ref{sec:methods} we present the models for a deformable prolate capsule and the membrane mechanics, and the numerical method adopted to calculate the fluid motion; in Section \ref{sec:results} we provide detailed results and discussion for the orbital dynamics of a prolate capsule over a broad range of various parameters, including membrane deformability, bending stiffness, initial orientation, aspect ratio of the capsule and the viscosity ratio between the inner and outer fluids. Predictions of the rheological properties for a dilute suspension of prolate capsules are also reported. Concluding remarks are presented in Section \ref{sec:conclusion}.

\section{MODEL FORMULATION} \label{sec:methods}
    \subsection{Model and discretization} \label{sec:model}

  \begin{figure}[h]
  \centering
  \captionsetup{justification=raggedright}
  \includegraphics[width=0.5\textwidth]{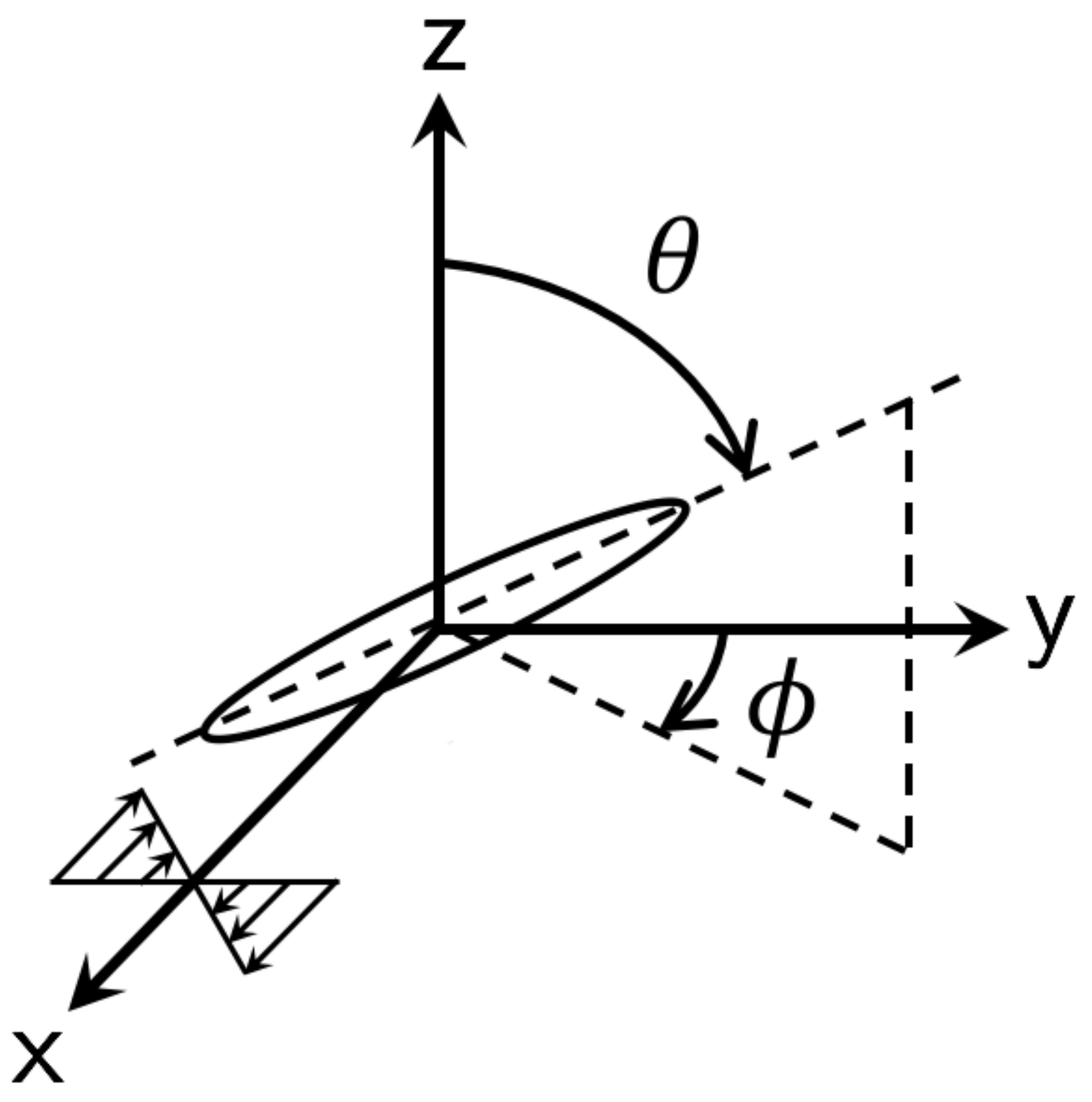}
  \caption[determination of orientation]{Schematic of the 3D orientation of a prolate capsule in unbounded simple shear flow.}
  \label{fig:orientation_determination}
  \end{figure}

We consider an isolated inertialess fluid-filled deformable capsule with a prolate spheroidal rest shape immersed in unbounded simple shear flow with shear rate $\dot{\gamma}$ (FIG.~\ref{fig:orientation_determination}). The undisturbed flow velocity is given by $\mathbf{u}^{\infty} = (\dot{\gamma} y, 0, 0)$. Both the suspending fluid and the fluid inside the capsule are assumed to be incompressible and Newtonian, with viscosity $\eta$ and $\lambda \eta$, respectively, where $\lambda$ is the viscosity ratio.  At rest, the prolate capsule has a polar radius $a_1$ and an equatorial radius $a_2$, with an aspect ratio AR = $a_1/a_2 > 1$. Here we define a characteristic length scale $a = (a_1 a_2^2)^{1/3}$; this is the radius of a sphere having the same volume as the prolate capsule. The particle Reynolds number $\textnormal{Re}_p = \rho \dot{\gamma} a^2/\eta$ is assumed to be sufficiently small so that the fluid motion is governed by the Stokes equation. The instantaneous orientation of the prolate capsule is given by two angles, as defined by Jeffery \cite{Jeffery:1922wb}: $\phi$ is the azimuthal angle with respect to the $y$ axis, and $\theta$ is the polar angle with respect to the $z$ axis.

To describe the membrane mechanics, we adopt a model that incorporates shear elasticity, area dilatation, and bending resistance. The total energy $E$ of the capsule membrane $S$ is given as:
\begin{equation} \label{eq:membrane_energy}
  E = \frac{K_B}{2} \int_S (2 \kappa_H + c_0)^2 dS + \overline{K_B} \int_S \kappa_G dS + \int_S W dS,
\end{equation}
where $K_B$ and $\overline{K_B}$ are the bending moduli, and $W$ is the shear strain energy density; $\kappa_H$ and $\kappa_G$ are the mean and Gaussian curvature of the membrane surface, respectively; $c_0 = -2H_0$ is the spontaneous curvature, $H_0$ being the mean curvature of the spontaneous shape. In this equation, the first two terms represent the Canham-Helfrich bending energy \cite{CANHAM197061,Helfrich1973}, and the third term corresponds to the shear strain energy. The behavior of the capsule membrane in response to the in-plane shear elastic force is described using a membrane model by Skalak \emph{et al.} \cite{SKALAK:1973tp}, in which the shear strain energy density $W$ is given by
\begin{equation} \label{eq:Skalak_model}
W_{\mathrm{SK}} = \frac{G}{4}[(I_1^2 + 2 I_1 - 2 I_2) + C_a I_2^2],
\end{equation}	
where $G$ is the in-plane shear modulus of the membrane, and $C_a$ characterizes the energy penalty for area change of the membrane. The strain invariants $I_1$ and $I_2$ are functions of the principal stretch ratios $\lambda_1$ and $\lambda_2$, defined as
\begin{equation} \label{eq:strain_invariants}
I_1 = \lambda_1^2 + \lambda_2^2 - 2, \quad I_2 = \lambda_1^2 \lambda_2^2 - 1.
\end{equation} 
Barth\`{e}s-Biesel \emph{et al.} \cite{DBB-diaz-dhenin-2002} showed that for $C_a \gtrsim 10$, the tension of a Skalak membrane becomes nearly independent of $C_a$ under a simple uniaxial deformation, so $C_a$ is set to $10$ for all of the simulations in the present work. The deformability and bending stiffness of the capsule are characterized by the dimensionless capillary number $\Ca$ = $ \eta \dot{\gamma} a/G$ and bending modulus $\hat{\kappa}_B = K_B/a^2 G$, respectively. Taking the first variation of the total membrane energy $E$ in Eq.~\ref{eq:membrane_energy} gives the total membrane strain force density:
\begin{equation} \label{eq:membrane_force}
\mathbf{f^m} = \mathbf{f^b} + \mathbf{f^s}, 
\end{equation}
where $\mathbf{f^b}$ and $\mathbf{f^s}$ are bending and shear elastic force densities, respectively. 

The capsule membrane is discretized into 320 piecewise flat triangular elements, resulting in 162 nodes. We have verified that increasing the number of nodes makes no difference to the cell dynamics. Based on this discretization, the calculation of the total membrane force density $\mathbf{f^m}$ follows the work of Kumar and Graham \cite{Kumar:2012ev} and Sinha and Graham \cite{Sinha:2015wt} using approaches given by Charrier \emph{et al.} \cite{charrier1989free} for the in-plane shear force density $\mathbf{f^s}$ and Meyer \emph{et al.} \cite{Meyer:2002vh} for the out-of-plane bending force density $\mathbf{f^b}$, respectively. Details regarding these calculations are found in \cite{Kumar:2012ev} and \cite{Sinha:2015wt}.   

    \subsection{fluid motion} \label{fluid_motion}
    
In the Stokes flow regime, the fluid velocity $\mathbf{u}$ at any point $\mathbf{x_0}$ in the unbounded domain can be written in boundary integral form \cite{pozrikidis1992boundary,Kumar:2012ev} as:
\begin{equation} \label{eq:BI_equation}
  u_j(\mathbf{x_0}) = u_j^{\infty}(\mathbf{x_0}) + \int_{S} q_i(\mathbf{x}) G_{ji}(\mathbf{x_0}, \mathbf{x}) dS(\mathbf{x}),
\end{equation}
where $\mathbf{q(x_0)}$ is a single-layer density that satisfies 
\begin{equation} \label{eq:single_layer_density}
  q_j(\mathbf{x_0}) + \frac{\lambda - 1}{4 \pi (\lambda + 1)} n_k(\mathbf{x_0}) \int_{S} q_i(\mathbf{x}) T_{jik}(\mathbf{x_0}, \mathbf{x}) dS(\mathbf{x}) = - \frac{1}{4 \pi \mu} \big(\frac{\Delta f_j(\mathbf{x_0})}{\lambda + 1} + \frac{\lambda - 1}{\lambda + 1} f_j^{\infty}(\mathbf{x_0})\big). 
\end{equation} 
Here $\mathbf{u}^{\infty}(\mathbf{x_0})$ is the undisturbed fluid velocity at a given point $\mathbf{x_0}$, $S$ denotes the surface of the particle; $\mathbf{f}^{\infty}(\mathbf{x_0})$ is the traction at a given point due to the stress generated in the fluid corresponding to the undisturbed flow $\mathbf{u}^{\infty}(\mathbf{x_0})$; $\Delta \mathbf{f}(\mathbf{x})$ is the hydrodynamic traction jump across the membrane interface, which relates to the total membrane force density by $\Delta \mathbf{f}(\mathbf{x}) = -\mathbf{f^m}$ assuming the membrane equilibrium condition; $\mathbf{G}$ and $\mathbf{T}$ are the Green's function and its associated stress tensor for Stokes flow in an unbounded domain. Details of the numerical method are described in \cite{Kumar:2012ev}. Once the flow field is determined, the positions of the element nodes on the discretized capsule membrane are advanced in time using a second-order Adams-Bashforth method with adaptive time step $\Delta t = 0.02$Ca$l$, where $l$ is the minimum node-to-node distance. Time is nondimensionalized by the shear rate $\dot{\gamma}$, and in this work we define $t$ as the dimensionless time. 

\XZrevise{The implementation of the boundary integral method used in this work has been extensively validated with test problems considering both a rigid sphere and a spherical drop between parallel walls subjected to pressure driven flow, and the numerical algorithms applied to calculate the shear and bending elastic force densities in the capsule membrane have been validated on the deformation of spherical capsules in simple shear flow. We have also compared the numerically determined inclination profile for a highly stiff prolate capsule with the prediction by Jeffery's theory \cite{Jeffery:1922wb} for an inertialess rigid prolate particle in shear flow, and observed good agreement. See \cite{Kumar:2012ev}, \cite{Sinha:2015wt} and \cite{Zhang2019} for details.}

    \subsection{Particle stress in a dilute suspension of capsules} \label{sec:rheology_theory}
    
\XZrevise{In addition to the capsule motions, we also compute the particle stress tensor in order to understand the rheology for a suspension of such capsules in the dilute limit.} For a suspension of $N_p$ capsules in a volume $V$, the contribution of the suspended capsules to the bulk stress of the suspension, in dimensional form, is given by \cite{KENNEDY1994251,Ramanujan:1998tx,Bagchi2010,graham_2018}
\begin{equation} \label{eq:dimensional_particle_stress}
  \Sigma_{ij}^p = \frac{1}{V} \sum\limits_{m=1}^{N_p}\int_{S_m} [\Delta f_i x_j + \eta (\lambda - 1) (u_i n_j + u_j n_i)] dS,
\end{equation}
where $\bm{\Sigma^p}$ is the particle stress tensor. \XZrevise{Rewritten in nondimensional variables,} $\widetilde{\textbf{x}} = \textbf{x}/a$, $\widetilde{\textbf{u}} = \textbf{u}/(a \dot \gamma)$, $\widetilde{V} = V/a^3$, $\widetilde{S} = S/a^2$, and $\widetilde{\Delta \textbf{f}} = \Delta \textbf{f}/(\eta \dot \gamma)$, Eq.~\ref{eq:dimensional_particle_stress} \XZrevise{becomes}
\begin{equation} \label{eq:dimensional_particle_stress_2}
  \Sigma_{ij}^p = \frac{\eta \dot \gamma}{\widetilde{V}} \sum\limits_{m=1}^{N_p}\int_{\widetilde{S}_m} [\widetilde{\Delta f}_i \widetilde{x}_j + (\lambda - 1) (\widetilde{u}_i n_j + \widetilde{u}_j n_i)] d\widetilde{S}.
\end{equation}
Now we assume that the suspension is dilute, containing $N_p$ identical prolate capsules all undergoing the same stable orbital motion, and that the interparticle hydrodynamic interactions are negligible. The volume fraction of the suspension is given by $\Phi = N_p V_p/V = N_p \widetilde{V}_p/\widetilde{V}$, where $V_p$ is the volume of a single capsule and $\widetilde{V}_p = V_p/a^3$ is dimensionless. Eq.~\ref{eq:dimensional_particle_stress_2} now becomes
\begin{equation} \label{eq:dimensional_particle_stress_3}
\begin{split}
  \Sigma_{ij}^p & = \frac{\eta \dot \gamma N_p}{\widetilde{V}}\int_{\widetilde{S}_m} [\widetilde{\Delta f}_i \widetilde{x}_j + (\lambda - 1) (\widetilde{u}_i n_j + \widetilde{u}_j n_i)] d\widetilde{S} \\
  & = \frac{\eta \dot \gamma \Phi}{\widetilde{V}_p}\int_{\widetilde{S}_m} [\widetilde{\Delta f}_i \widetilde{x}_j + (\lambda - 1) (\widetilde{u}_i n_j + \widetilde{u}_j n_i)] d\widetilde{S},
\end{split}
\end{equation}
which naturally gives the dimensionless particle stress tensor
\begin{equation} \label{eq:dimensionless_particle_stress}
  \widetilde{\Sigma^p_{ij}} = \frac{\Sigma_{ij}^p}{\eta \dot \gamma \Phi} = \frac{1}{\widetilde{V}_p}\int_{\widetilde{S}_m} [\widetilde{\Delta f}_i \widetilde{x}_j + (\lambda - 1) (\widetilde{u}_i n_j + \widetilde{u}_j n_i)] d\widetilde{S}.
\end{equation}
The dimensionless particle shear stress $\widetilde{\Sigma^p_{xy}}$ is \XZrevise{identical to} the intrinsic viscosity $[\eta]$; for a dilute suspension of rigid spherical particles \XZrevise{$[\eta] = 2.5$} \cite{Einstein1911}. The dimensionless first and second normal stress differences \XZrevise{are} $N_1 = \widetilde{\Sigma^p_{xx}} - \widetilde{\Sigma^p_{yy}}$ and $N_2 = \widetilde{\Sigma^p_{yy}} - \widetilde{\Sigma^p_{zz}}$.
      
\section{RESULTS AND DISCUSSION} \label{sec:results}

\XZrevise{In this section, the orbital dynamics of a prolate capsule in unbounded simple shear flow are systematically investigated over a broad domain of parameter space. We focus on addressing the issue of \XZrevise{long-time behavior} of the orbits. For a capsule with small bending stiffness (Section~\ref{sec:low_bending}), \XZrevise{all initial conditions we consider evolve towards the same orbit at long times}. For a capsule with large bending stiffness (Section~\ref{sec:high_bending}), in contrast, we reveal \XZrevise{parameter regimes with multiple attractors}, i.e., \XZrevise{there are multiple stable orbits, and which one is reached at long times depends on the initial orientation}. A corresponding multiplicity in the rheological properties is predicted for a dilute suspension of such capsules.}
 
   \subsection{Dynamics of prolate capsules with small bending stiffness} 
      \label{sec:low_bending}
      
\begin{figure}[t]
\centering
\captionsetup{justification=raggedright}
 \subfloat[]
{
    \includegraphics[width=0.4\textwidth]{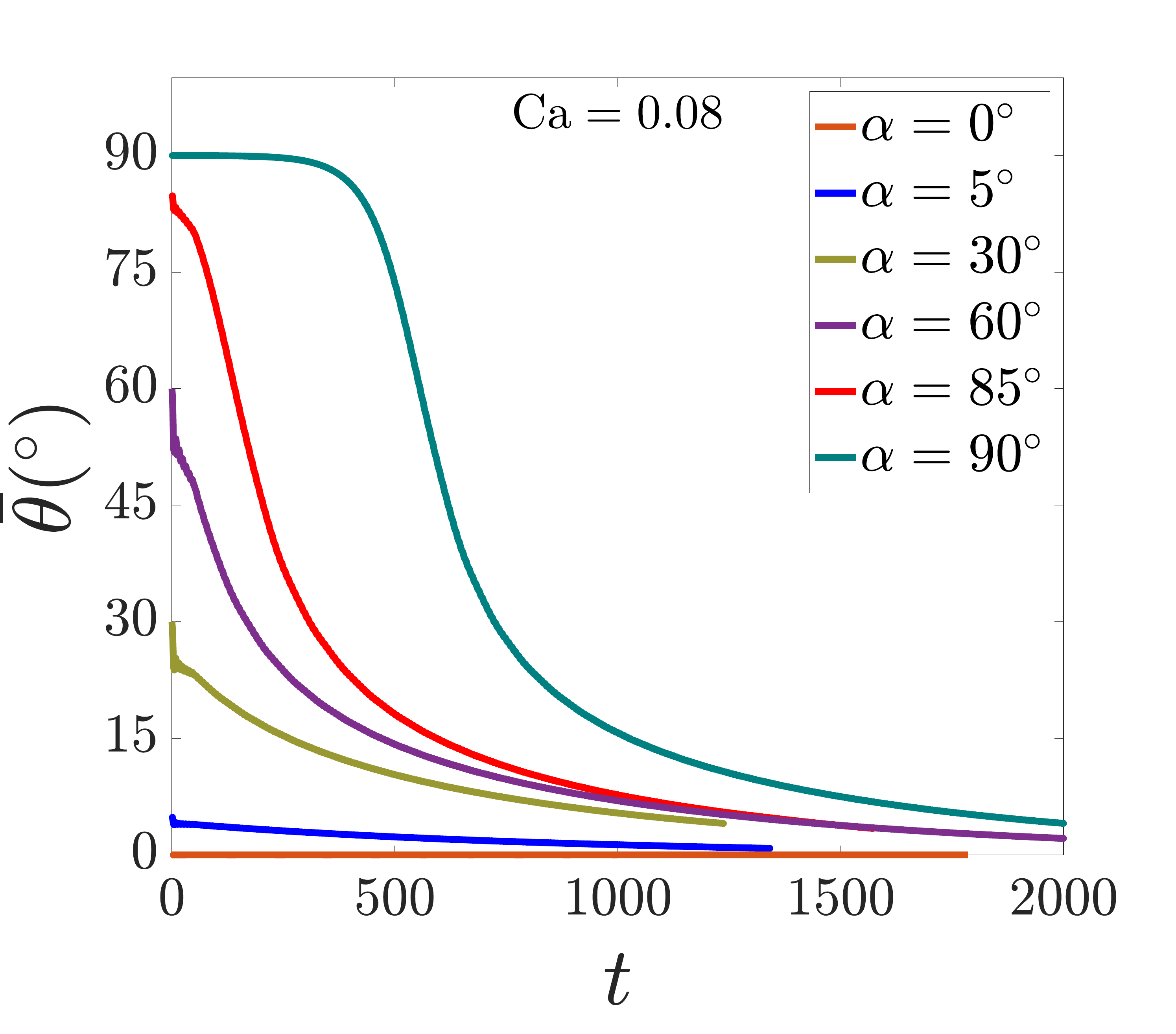}
    \label{fig:rolling}
}
 \subfloat[]
{
    \includegraphics[width=0.4\textwidth]{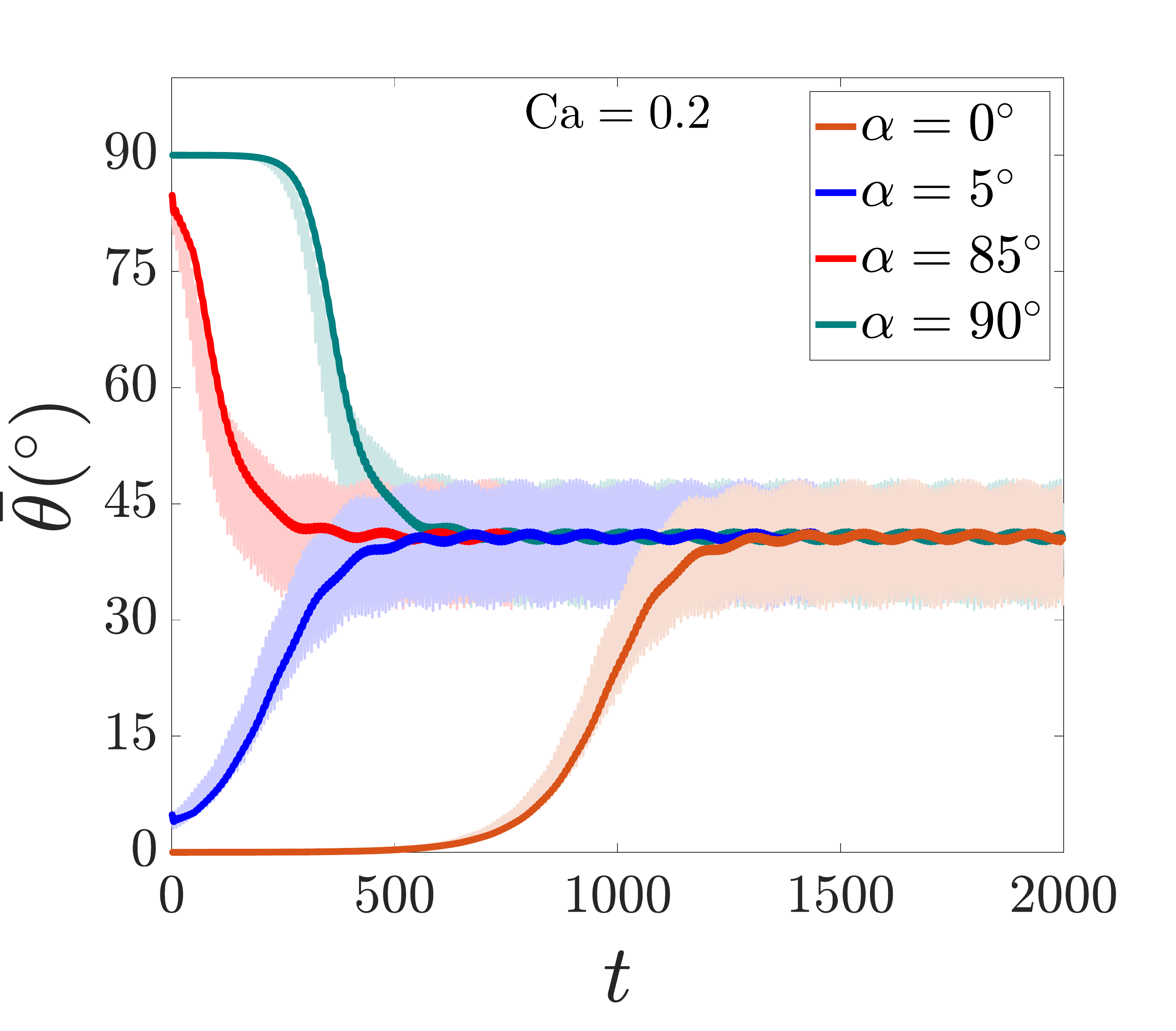}
    \label{fig:wobbling}
}
\\
 \subfloat[]
{
    \includegraphics[width=0.4\textwidth]{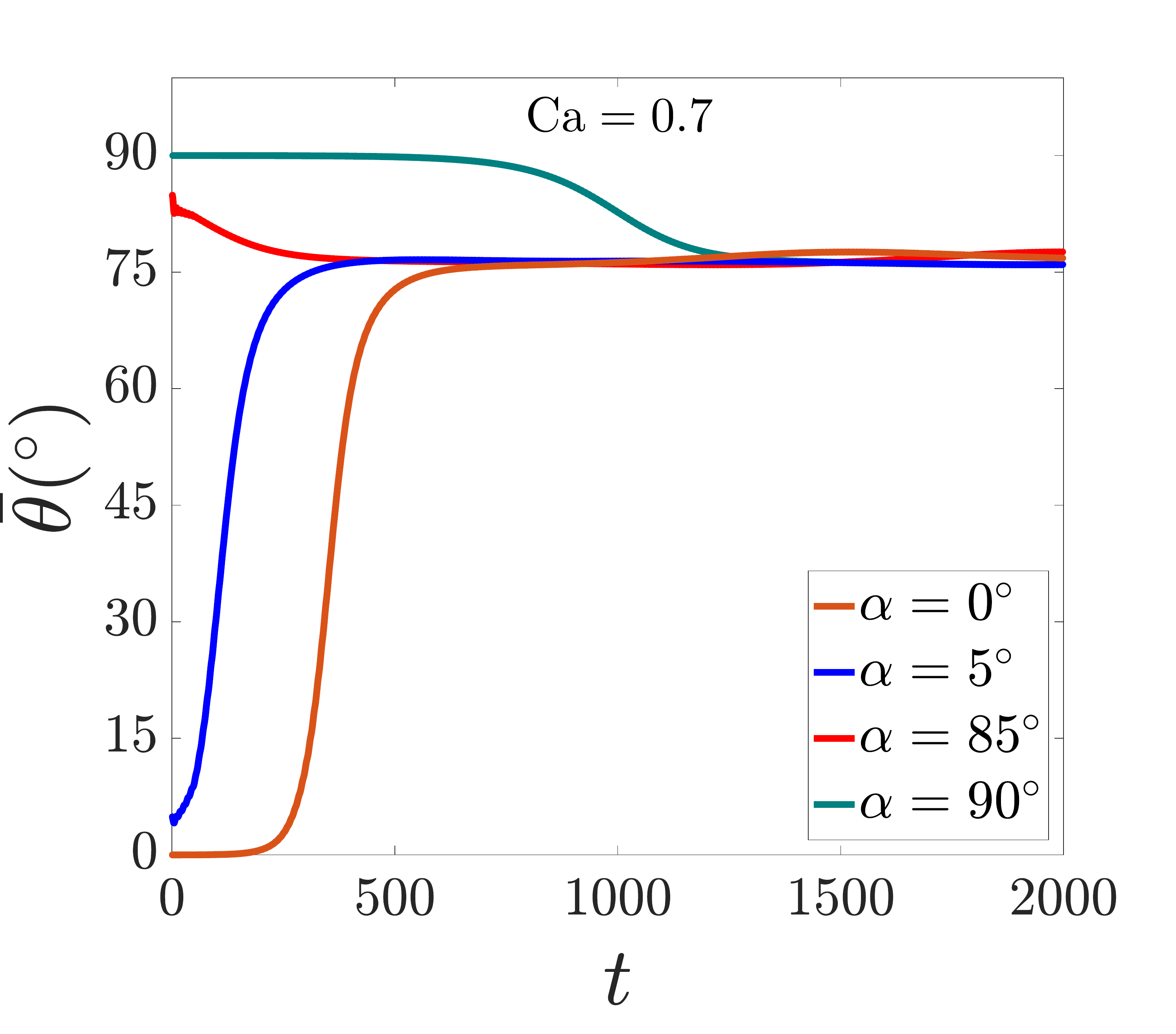}
    \label{fig:oscillating_swinging}
}
 \subfloat[]
{
    \includegraphics[width=0.4\textwidth]{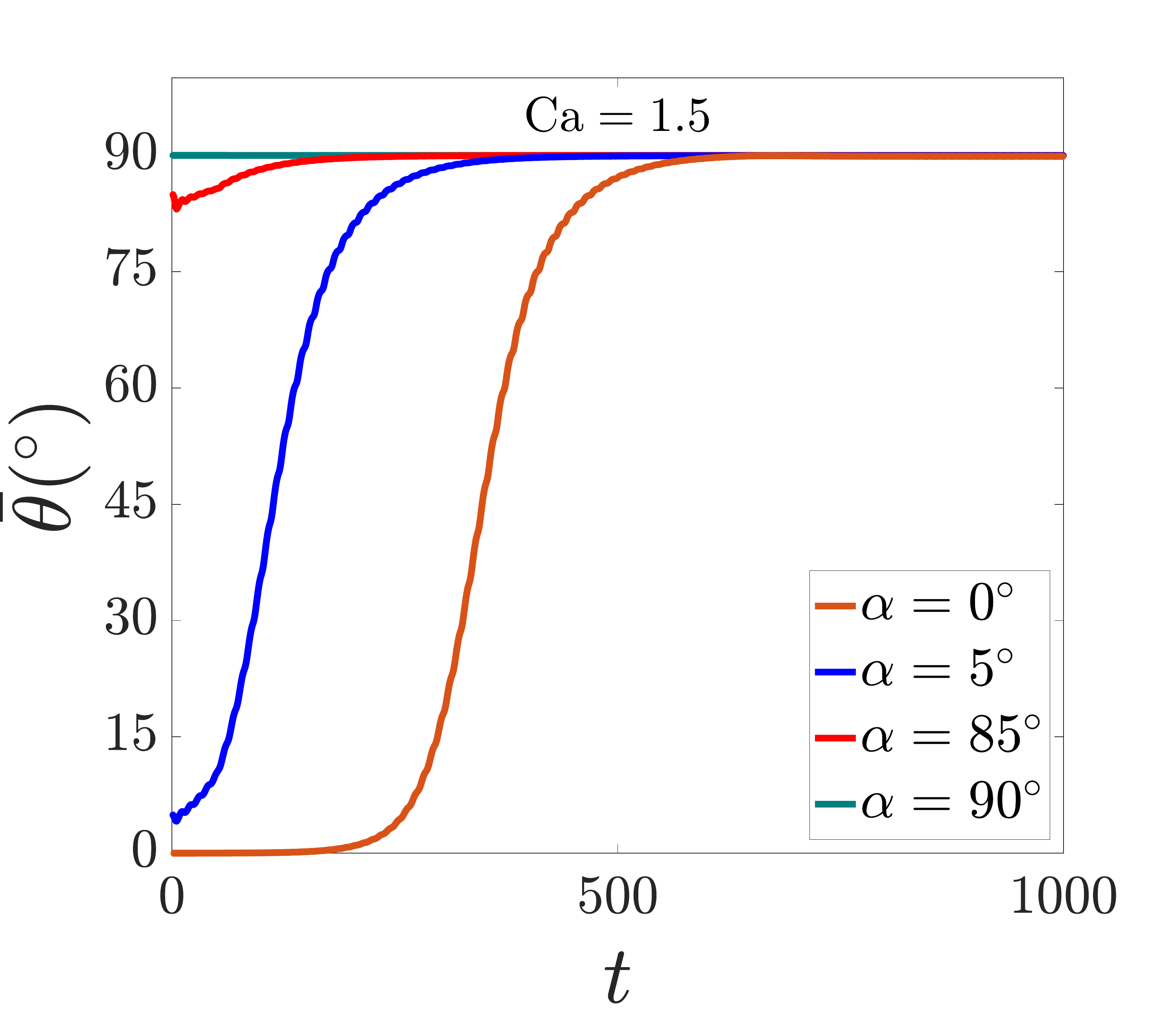}
    \label{fig:pure_swinging}
}

\caption{Evolution of $\bar \theta$ of a prolate capsule (AR = 2.0) with $\hat{\kappa}_B = 0.02$ and $\lambda = 1$ at (a) $\Ca$ = 0.08, (b) $\Ca$ = 0.2, (c) $\Ca$ = 0.7, and (d) $\Ca$ = 1.5. Here the spontaneous shape of the capsule is assumed to be the same as its rest shape.}
\label{lambda_1_small_bending}
\end{figure}

We \XZrevise{first} report the dynamics of a deformable prolate capsule with small bending stiffness ($\hat{\kappa}_B = 0.02$). The initial orientation of the capsule is set to $[\phi^0,\theta^0] = [\pi/2, \alpha]$. We \XZrevise{begin with} the case where $\lambda = 1$ and AR = 2.0\XZrevise{, and assume that the spontaneous shape of the capsule is the same as its rest shape. The effect of spontaneous curvature will be discussed later.} FIG.~\ref{lambda_1_small_bending} shows the evolution of the capsule \XZrevise{orbit} in various regimes of $\Ca$. We are interested in the long-time limit (dimensionless time $t \gg 1$) of the capsule dynamics, so the evolution data is presented using a running average of $\theta$, defined as $\bar \theta = \frac{1}{t_{\mathrm{avg}}} \int_{t-t_{\mathrm{avg}}}^{t} \theta(t) dt$ that averages over the oscillations in $\theta$ caused by the rotational motions of the capsule on the time scale $1/\dot{\gamma}$. Here we set $t_{\mathrm{avg}} = 50$; \XZrevise{FIG.~\ref{fig:wobbling} shows an example of the evolution data before (light-colored ``clouds") and after (solid colored lines) taking the running average.} 
  
\begin{figure}[t]
  \centering
  \captionsetup{justification=raggedright}
  \includegraphics[width=0.7\textwidth]{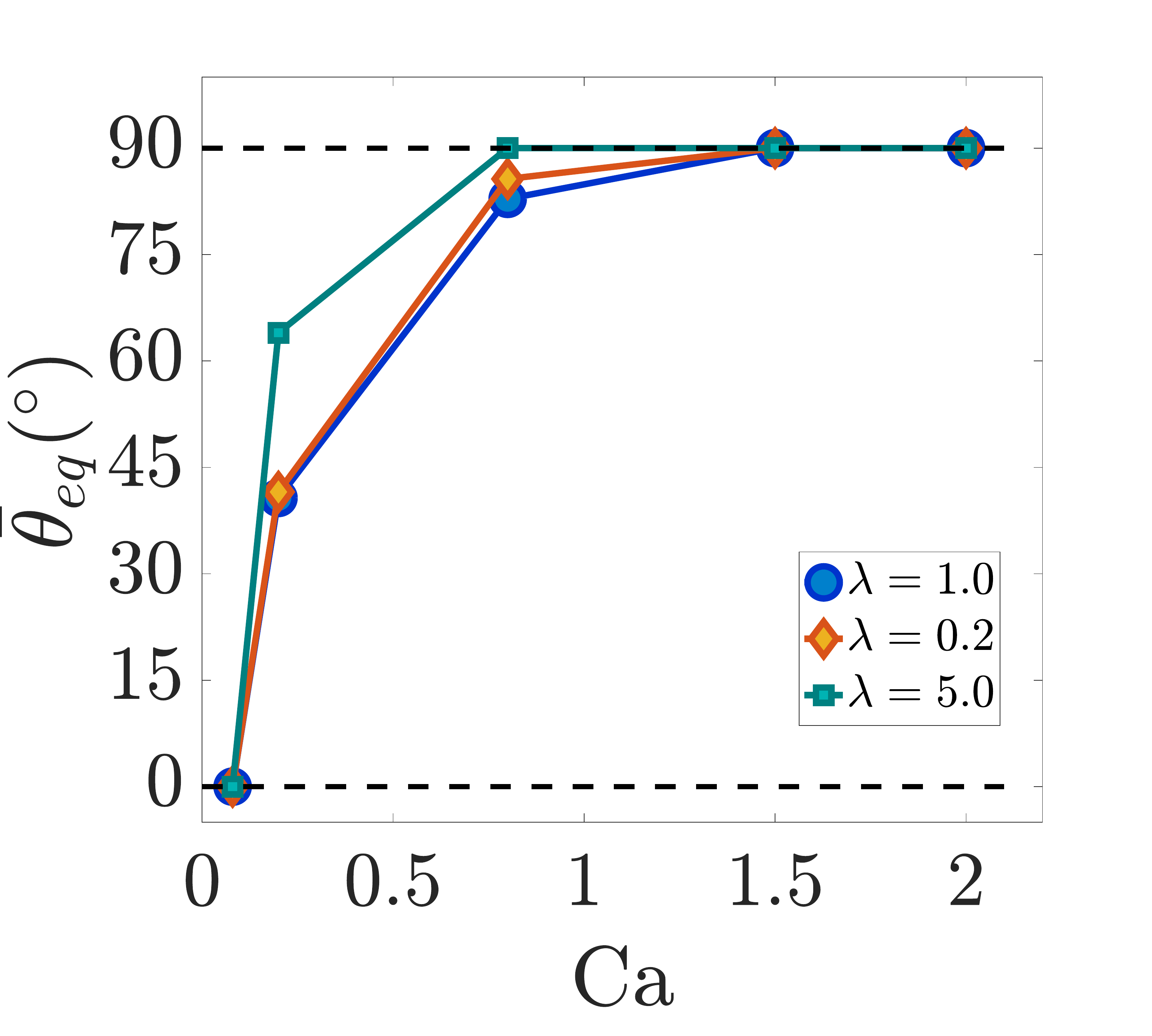}
  \caption{Bifurcation behavior for a prolate capsule (AR = 2.0) with $\hat{\kappa}_B = 0.02$ and varying $\lambda$. The black dashed lines at $\theta = 0^{\circ}$ and $\theta = 90^{\circ}$ represent the orbits for a capsule aligned with the $z$ axis and in the shear plane, respectively. }
  \label{fig:bifurcation_low_bending}
  \end{figure}

The key observation here is that the capsule, irrespective of the initial orientation $\alpha$, evolves towards the same orbit at long times, denoted as $\bar \theta_{eq}$. \XZrevise{In FIG.~\ref{fig:rolling} ($\Ca = 0.08$), six values of $\alpha$ are considered, and the orbit of the capsule always converges towards $\bar \theta_{eq} = 0^{\circ}$. For simplicity, results are only shown for four values of $\alpha$ in FIGs.~\ref{fig:wobbling}, \ref{fig:oscillating_swinging} and \ref{fig:pure_swinging}}. As $\Ca$ increases, $\bar \theta_{eq}$ increases from $0^{\circ}$ to $90^{\circ}$. This trend holds qualitatively for $\lambda = 0.2$ and $\lambda = 5$, and also as AR increases from 2.0 to 3.0 (not shown). \XZrevise{The dependence of $\bar \theta_{eq}$ on $\Ca$ for a prolate capsule (AR = 2.0) with $\hat{\kappa}_B = 0.02$ is illustrated in a bifurcation diagram with varying $\lambda$, as shown in FIG.~\ref{fig:bifurcation_low_bending}. Note that by symmetry, the orbits for a capsule lying in the shear plane ($\theta = 90^{\circ}$) and aligned with the $z$ axis ($\theta = 0^{\circ}$) are always solutions for all parameter values. However, these two solutions may not be stable with respect to symmetry-breaking perturbations; indeed, in the case of $\lambda = 1$ (solid blue line with circles), for example, there is a symmetry-breaking bifurcation away from $\theta = 0^{\circ}$ between $\Ca = 0.08$ and $\Ca = 0.2$, and another one away from $\theta = 90^{\circ}$ between $\Ca = 0.7$ and $\Ca = 1.5$.}

\begin{figure}[H]
\centering
\captionsetup{justification=raggedright}
 \subfloat[]
{
    \includegraphics[width=0.65\textwidth]{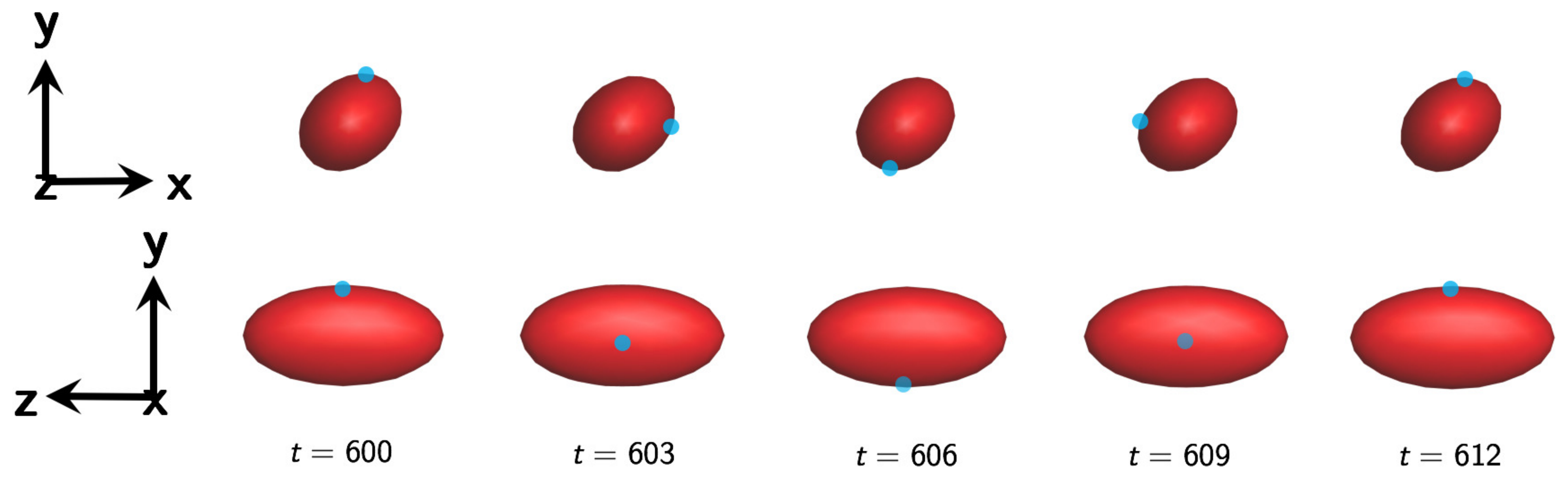}
    \label{fig:rolling_snapshots_1}
} 
\\
\subfloat[]
{
    \includegraphics[width=0.65\textwidth]{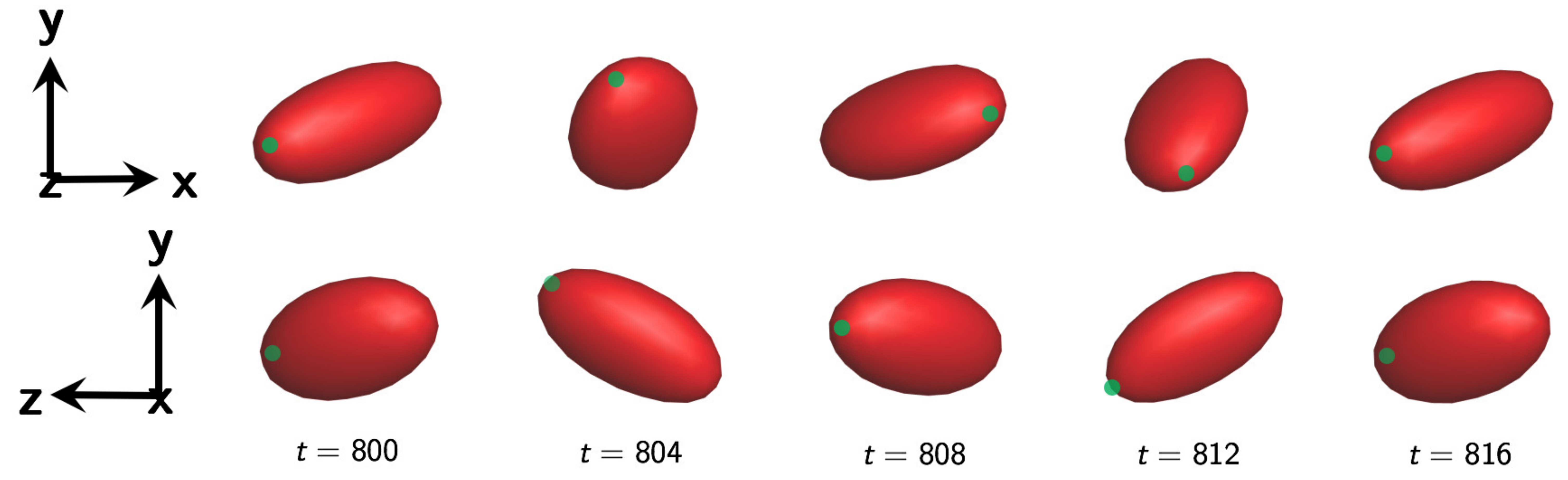}
    \label{fig:wobbling_snapshots_1}
}
\\
\subfloat[]
{
    \includegraphics[width=0.65\textwidth]{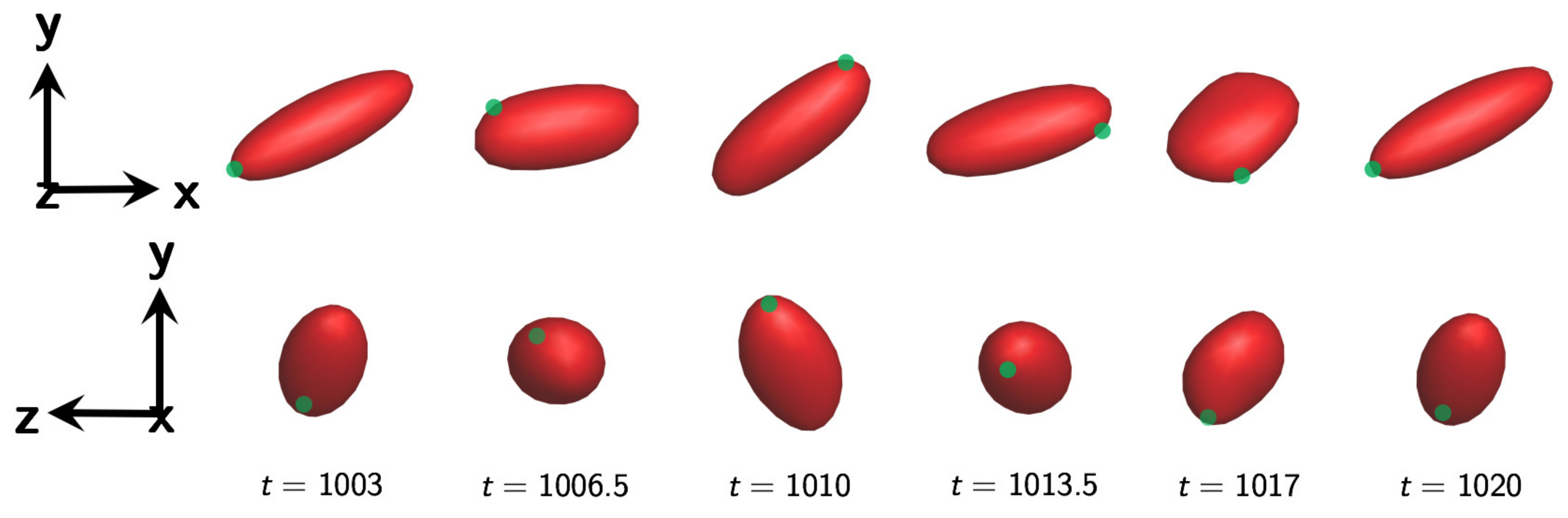}
    \label{fig:oscillating_swinging_snapshots_1}
}
\\
\subfloat[]
{
    \includegraphics[width=0.65\textwidth]{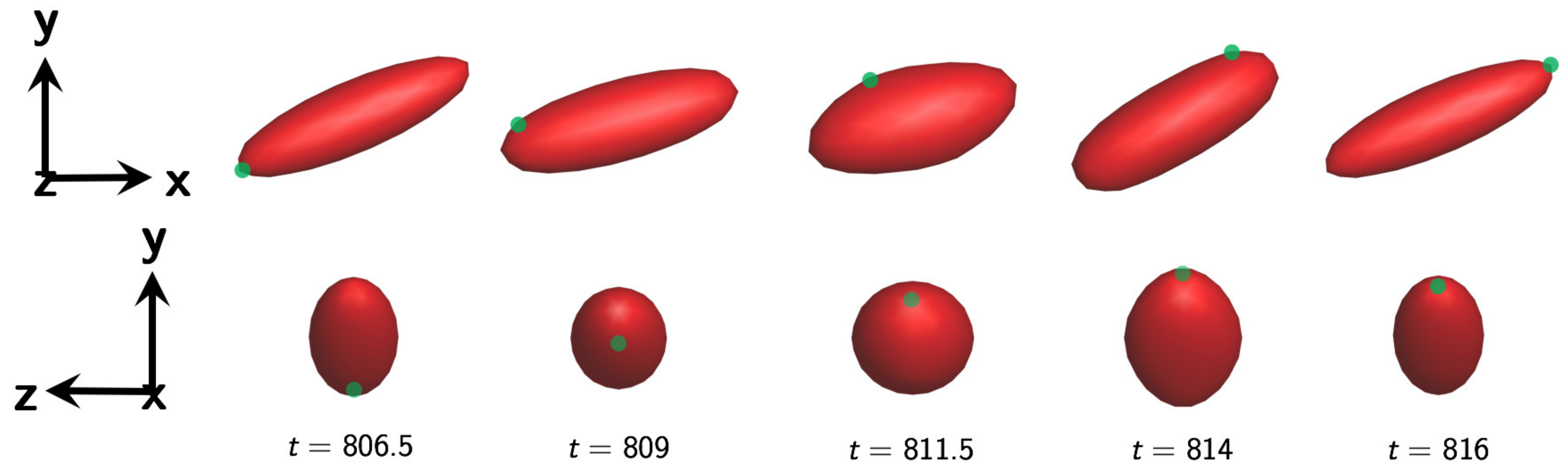}
    \label{fig:swinging_snapshots_1}
}
\caption{Time sequence images (side and front views) of a prolate capsule (AR = 2.0) with $\hat{\kappa}_B = 0.02$ and $\lambda = 1$ taking a rolling (a, $\Ca$ = 0.08), wobbling (b, $\Ca$ = 0.2), oscillating-swinging (c, $\Ca$ = 0.7), and swinging (d, $\Ca$ = 1.5) motion, respectively. \XZrevise{The initial orientation of the capsule is $\alpha = 5^{\circ}$. The markers indicate the positions of a membrane point initially on the equator of the capsule for (a), and a membrane point initially on the major axis of the capsule for (b), (c) and (d).}} 
\label{snapshots_low_bending}
\end{figure}

\XZrevise{We now look into the detailed motions of the capsule in different $\Ca$ regimes. With increasing $\Ca$, four orbital modes are determined corresponding to the stable orbits in different regimes.} At very low $\Ca$, the major axis of the capsule is observed to evolve towards the vorticity axis via a process termed drifting precession \cite{Cordasco:2013hb}, till the capsule takes a stable rolling motion (FIG.~\ref{fig:rolling_snapshots_1}). As $\Ca$ increases, the capsule orbit drifts towards an intermediate equilibrium configuration where the capsule exhibits a wobbling behavior while precessing about the vorticity axis (FIG.~\ref{fig:wobbling_snapshots_1}); this motion is named stable precession in \cite{Cordasco:2013hb}. In the regime of higher $\Ca$, the capsule adopts a complex motion termed oscillating-swinging (FIG.~\ref{fig:oscillating_swinging_snapshots_1}), in which the long axis of the elongated capsule oscillates both about the shear plane and about a mean inclination with respect to the flow direction, while the membrane of the capsule rotates around its deformed shape with the capsule elongation oscillating over time, as described by Dupont \emph{et al.} \cite{Dupont2013}. With a further increase in $\Ca$ (FIG.~\ref{fig:swinging_snapshots_1}), the oscillations about the shear plane in the oscillating-swinging motion vanish, and a pure in-plane fluid-like swinging motion, as described by Walter \emph{et al.} \cite{walter-salsac-DBB-2011}, is assumed eventually. This qualitative transition of \XZrevise{stable} orbital motions for a prolate capsule agrees well with the findings by Dupont \emph{et al.} \cite{Dupont2013}, which did not include bending resistance. 

\begin{figure}[t]
\centering
\captionsetup{justification=raggedright}
 \subfloat[]
{
    \includegraphics[width=0.4\textwidth]{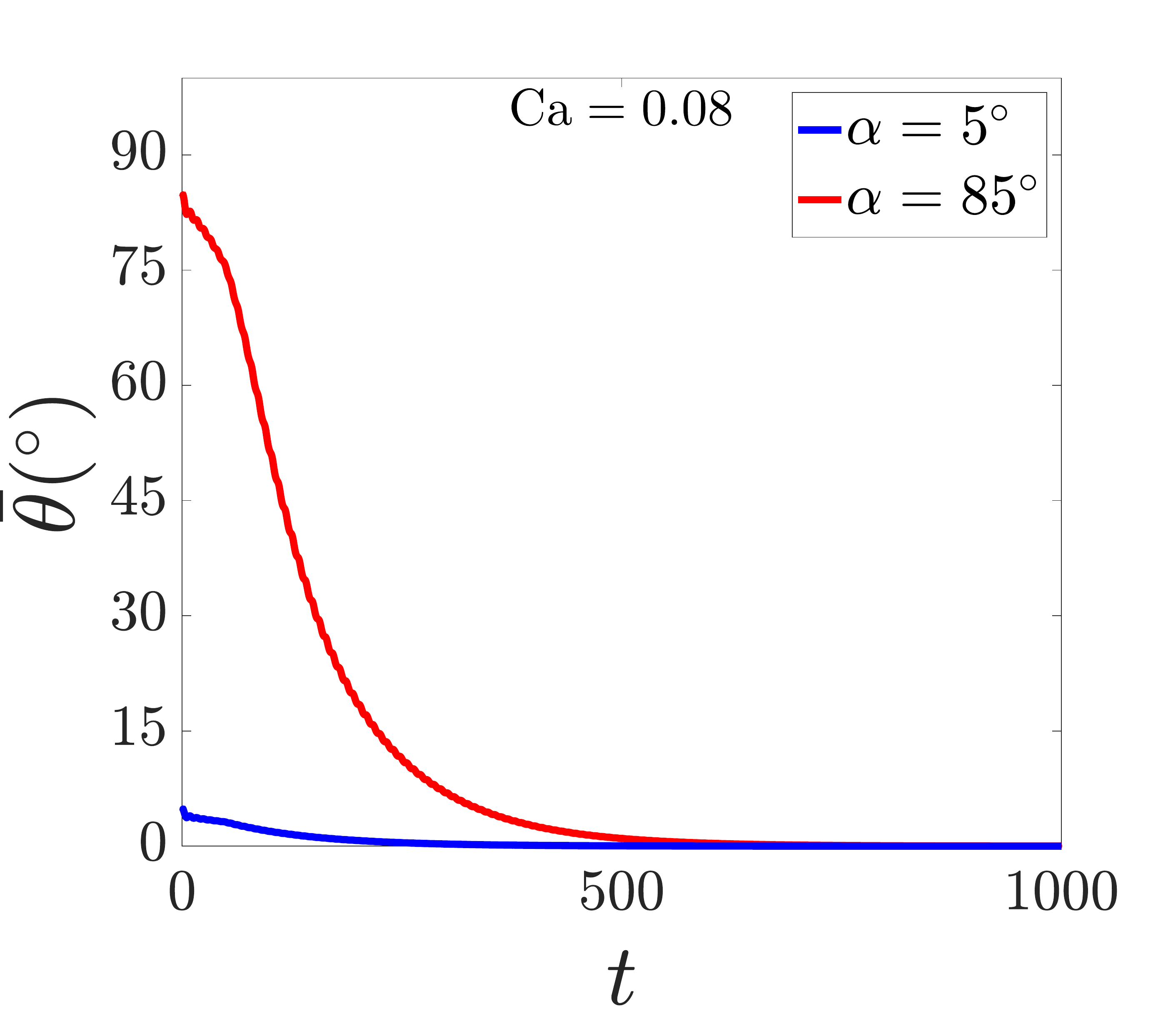}
}
 \subfloat[]
{
    \includegraphics[width=0.4\textwidth]{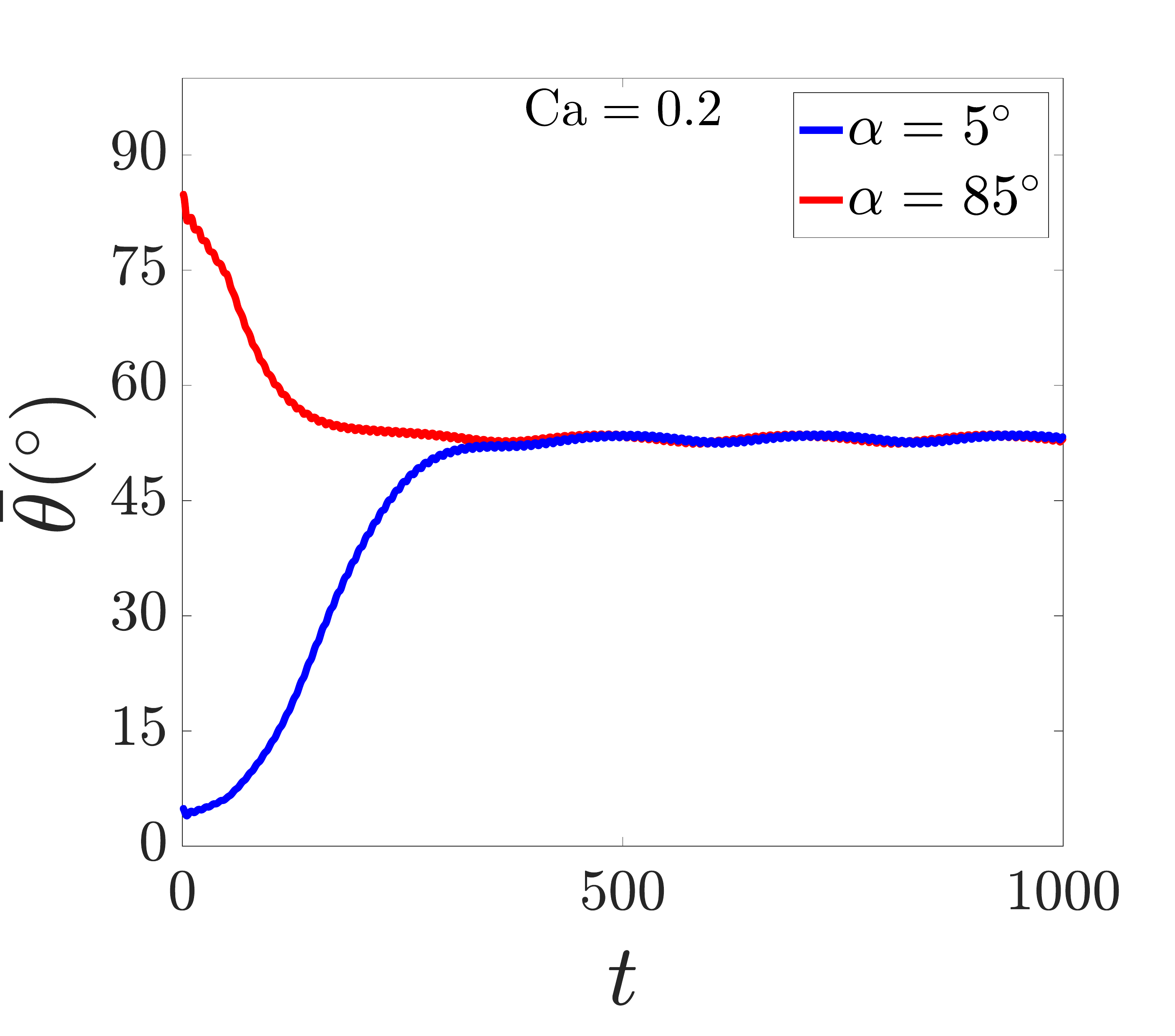}
}
\\
 \subfloat[]
{
    \includegraphics[width=0.4\textwidth]{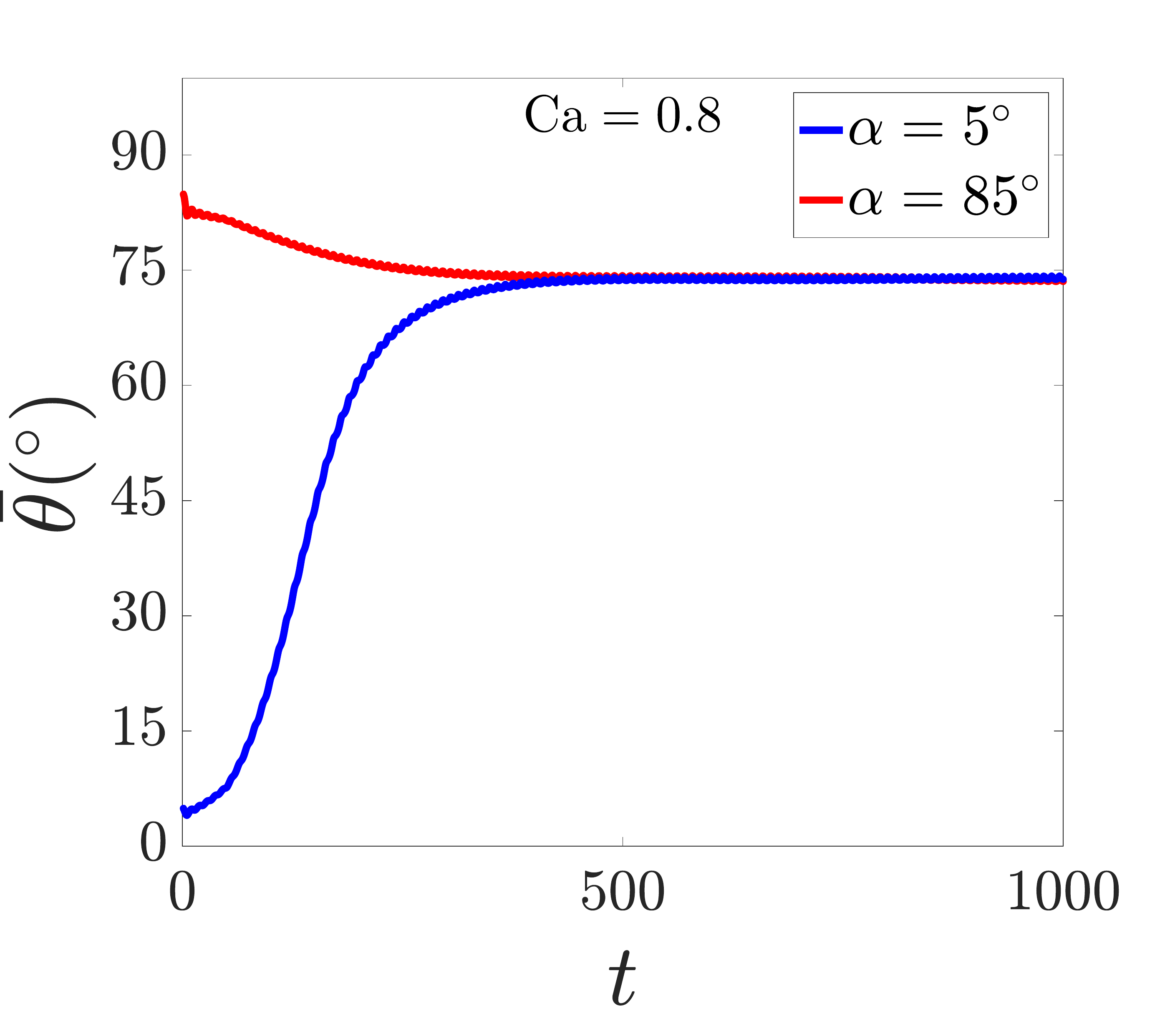}
}
 \subfloat[]
{
    \includegraphics[width=0.4\textwidth]{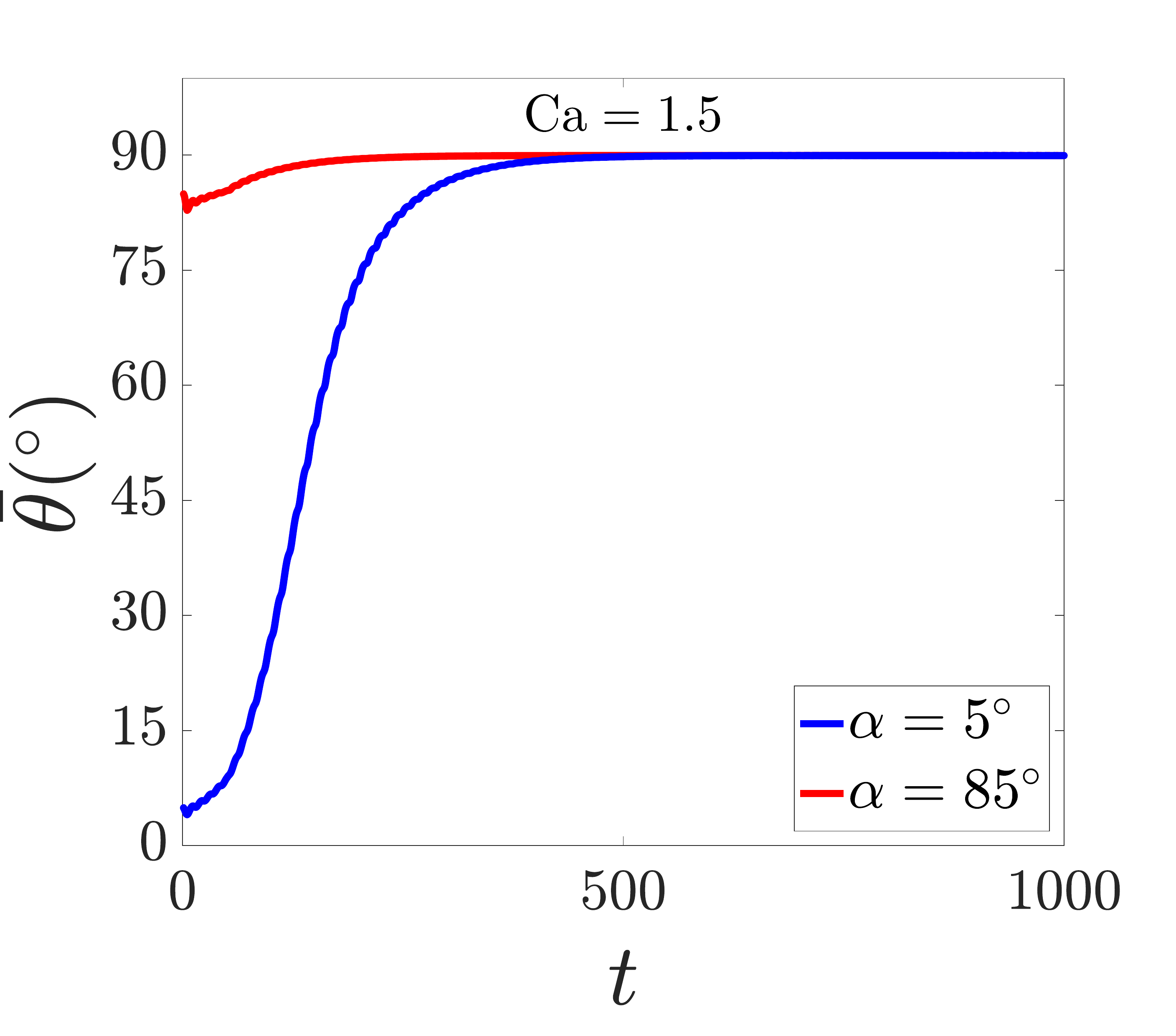}
}

\caption{Evolution of $\bar \theta$ of a prolate capsule (AR = 2.0) with $\hat{\kappa}_B = 0.02$ and $\lambda = 1$ at (a) $\Ca$ = 0.08, (b) $\Ca$ = 0.2, (c) $\Ca$ = 0.8, and (d) $\Ca$ = 1.5. Here the spontaneous curvature is assumed to be zero ($c_0 = 0$) everywhere on the membrane surface.}
\label{lambda_1_small_bending_zero_curvature}
\end{figure}

In addition, we examine the effect of spontaneous curvature on the orbital dynamics of the prolate capsule. FIG.~\ref{lambda_1_small_bending_zero_curvature} shows the orbit evolution for a capsule with zero spontaneous curvature everywhere on the membrane surface. \XZrevise{For each $\Ca$, the long-time orbit is independent of the initial orientation, and thus only the results for $\alpha = 5^{\circ}$ and $\alpha = 85^{\circ}$ are shown here.} Similar to the previous case, $\bar \theta_{eq}$ is observed to increase from $0^{\circ}$ to $90^{\circ}$ at increasing $\Ca$. Same stable motion modes are determined, namely, rolling, wobbling, oscillating-swinging, and swinging, associated with different $\Ca$ regimes. Cordasco and Bagchi \cite{Cordasco:2013hb} also found a qualitatively similar transition of the orbital dynamics at increasing $\Ca$ from drift precession (the rolling regime) to stable precession accompanied by a tank-treading behavior of the membrane (the oscillating-swinging regime) for a prolate capsule (AR = 2.0, $\lambda = 1$) with zero spontaneous curvature and small bending stiffness ($\hat{\kappa}_B = 0.01$). These results suggest that the spontaneous curvature has a minor effect on the orbital behavior of the capsule. 

Using Eq.~\ref{eq:dimensionless_particle_stress}, we can predict the rheology for a suspension of prolate capsules in the dilute limit. FIG.~\ref{rheology_small_bending} shows \MDGrevise{the capillary-number dependence of} the time-averaged intrinsic viscosity and the dimensionless normal stress differences predicted for a dilute suspension of such capsules \MDGrevise{all of which assume that same orbit} for AR = 2.0 and $\hat{\kappa}_B = 0.02$. A decrease in the intrinsic viscosity $[\eta]$ with increasing $\Ca$, i.e.~shear-thinning, is observed for all viscosity ratios considered here, and $[\eta]$ decreases with increasing $\lambda$ at each $\Ca$ (FIG.~\ref{fig:Sigma_xy_small_bending}). \MDGrevise{Shear-thinning is the generally-expected behavior for dilute suspensions of deformable particles, and indeed has also}{ been observed for dilute suspensions of other types of capsules, such as spherical capsules \cite{BARTHESBIESEL1981493,pozrikidis_1995,Ramanujan:1998tx,Bagchi2010,Clausen2010}, oblate capsules \cite{bagchi_kalluri_2011}, and biconcave discoidal capsules \cite{Pozrikidis2003,Drochon2003,takeishi_rosti_imai_wada_brandt_2019}. The dimensionless normal stress differences $N_1$ and $N_2$ are shown in FIGs.~\ref{fig:N1_small_bending} and \ref{fig:N2_small_bending}, respectively. For all cases, $N_1 > 0$ and $N_2 < 0$, with $|N_2|\ll N_1$. In general, both $N_1$ and $|N_2|$ decrease with increasing $\lambda$ at each $\Ca$, and increase monotonically with $\Ca$ for $\lambda = 1.0$ and $\lambda = 0.2$. For $\lambda = 5.0$, however, $N_1$ is observed to show a non-monotonic dependency on $\Ca$, while $|N_2|$ decreases with $\Ca$. 

\begin{figure}[h]
\centering
\captionsetup{justification=raggedright}
 \subfloat[]
{
    \includegraphics[width=0.45\textwidth]{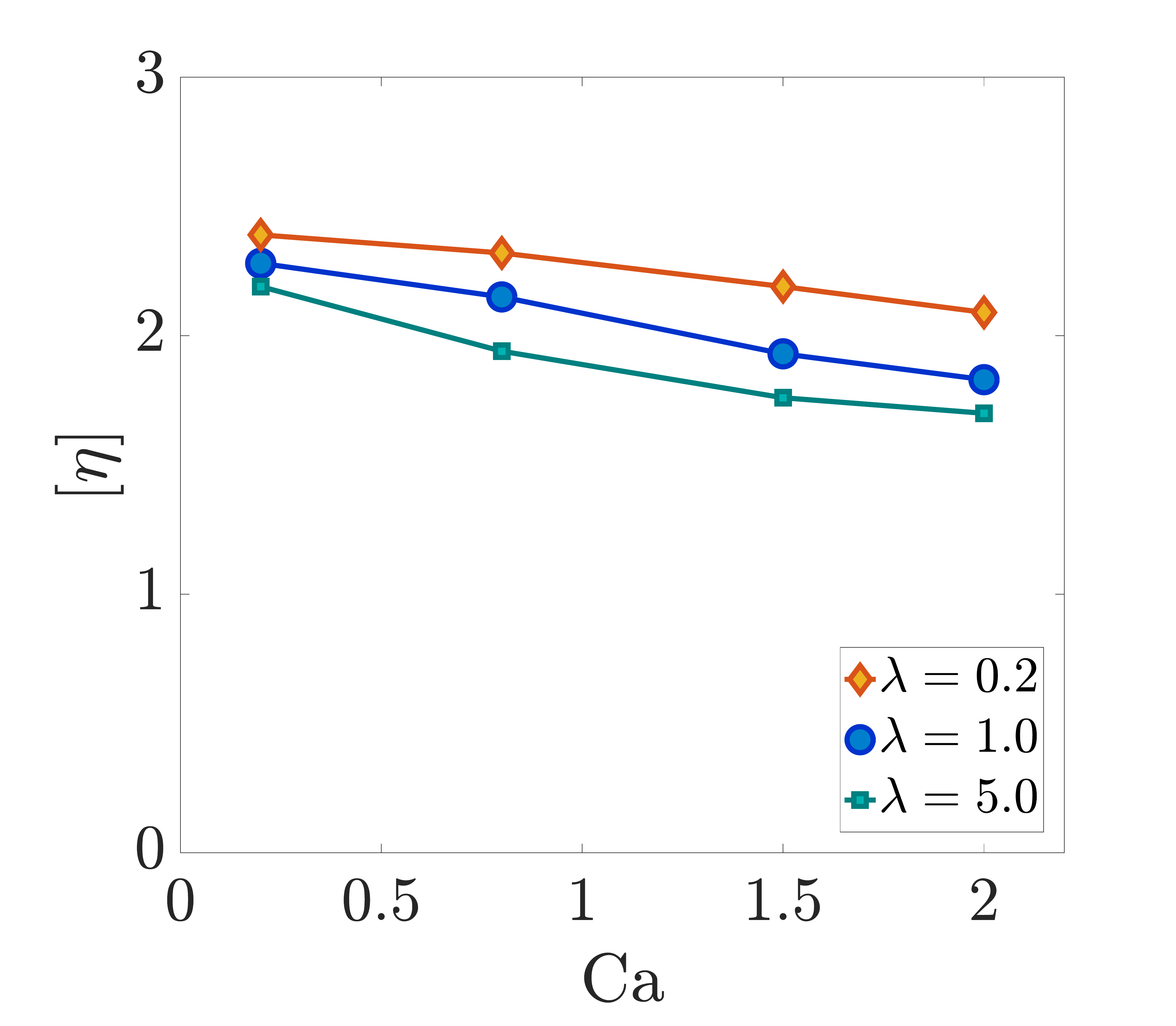}
    \label{fig:Sigma_xy_small_bending}
}
\\
 \subfloat[]
{
    \includegraphics[width=0.45\textwidth]{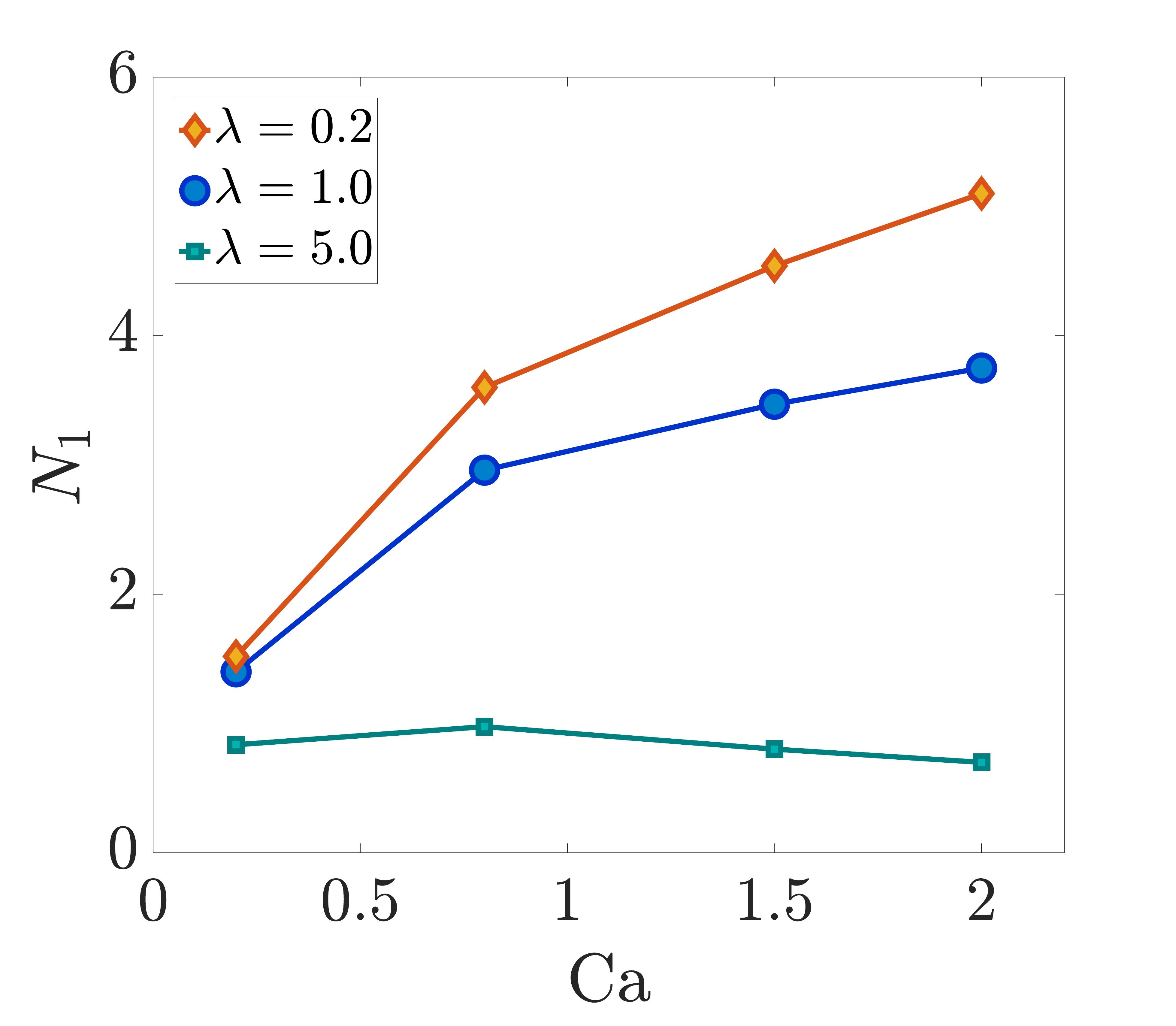}
    \label{fig:N1_small_bending}
}
 \subfloat[]
{
    \includegraphics[width=0.45\textwidth]{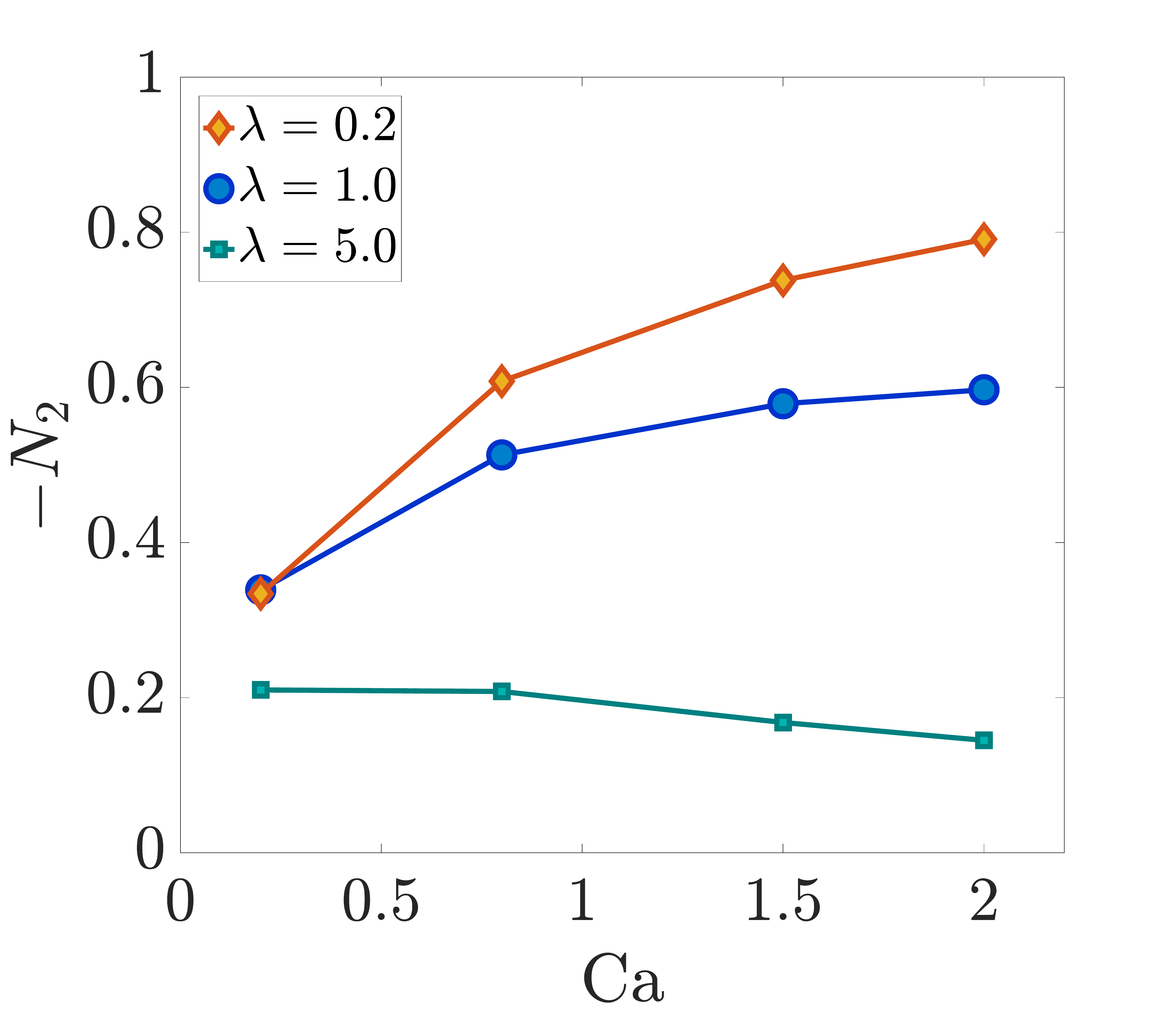}
    \label{fig:N2_small_bending}
}

\caption{Intrinsic viscosity (a) and dimensionless normal stress differences (b,c) predicted for a dilute suspension of identical prolate capsules (AR = 2.0) with $\hat{\kappa}_B = 0.02$ taking the same corresponding stable orbit at varying $\Ca$.}
\label{rheology_small_bending}
\end{figure}

   \subsection{Dynamics of prolate capsules with large bending stiffness} 
      \label{sec:high_bending}

\begin{figure}[h]
\centering
\captionsetup{justification=raggedright}
 \subfloat[]
{
    \includegraphics[width=0.4\textwidth]{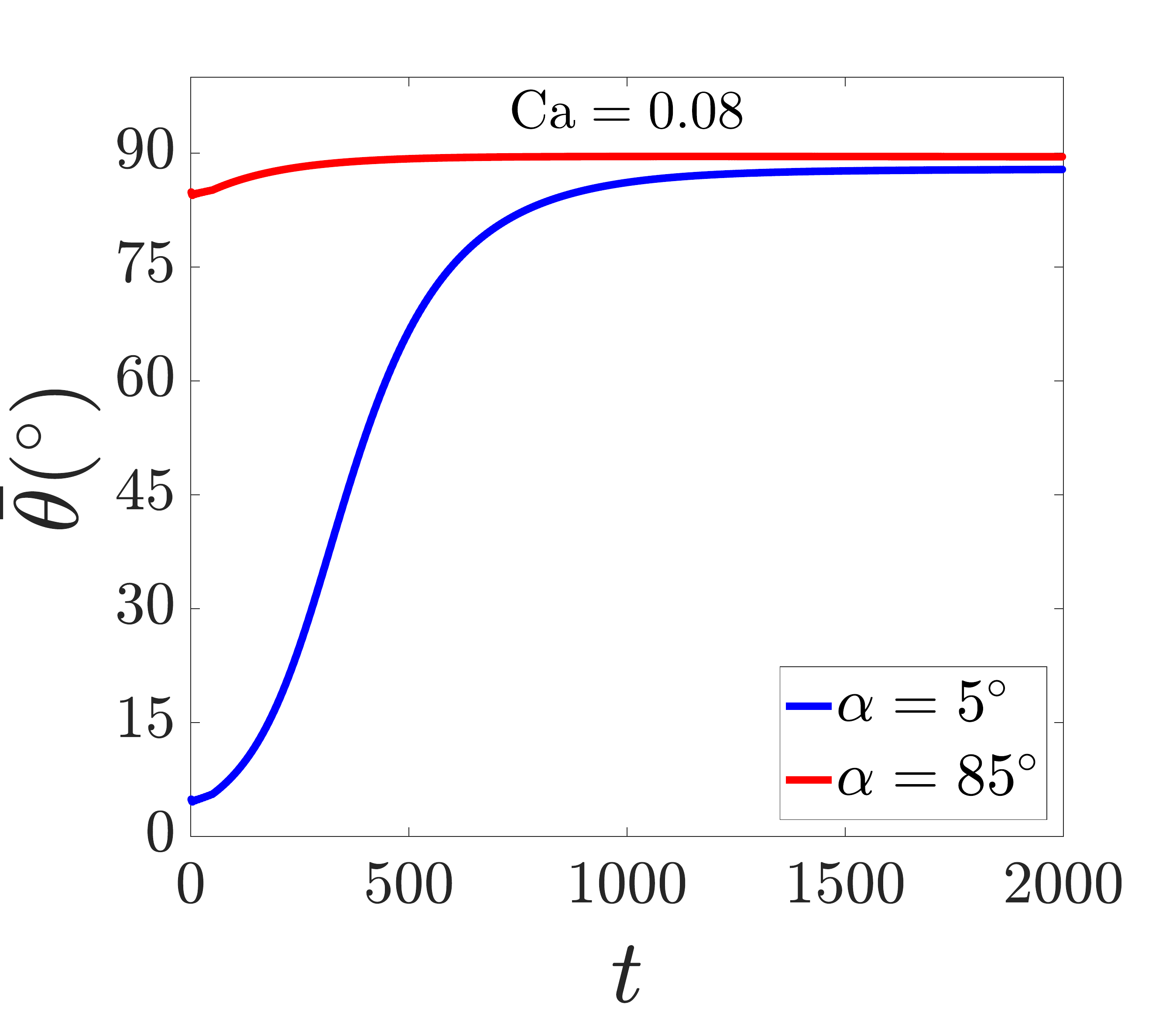}
    \label{fig:tumbling_1}
}
 \subfloat[]
{
    \includegraphics[width=0.4\textwidth]{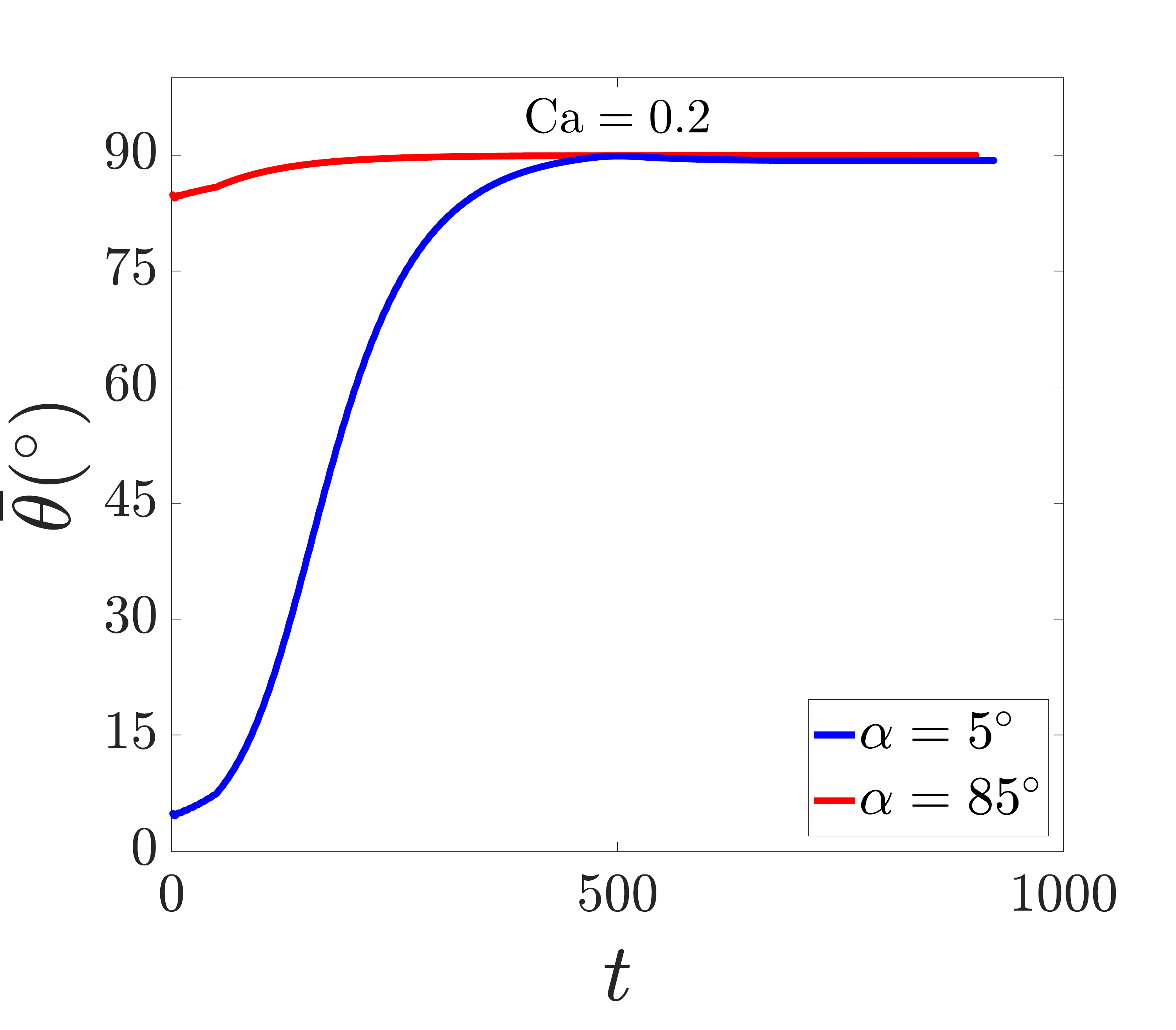}
    \label{fig:tumbling_2}
}
\\
 \subfloat[]
{
    \includegraphics[width=0.4\textwidth]{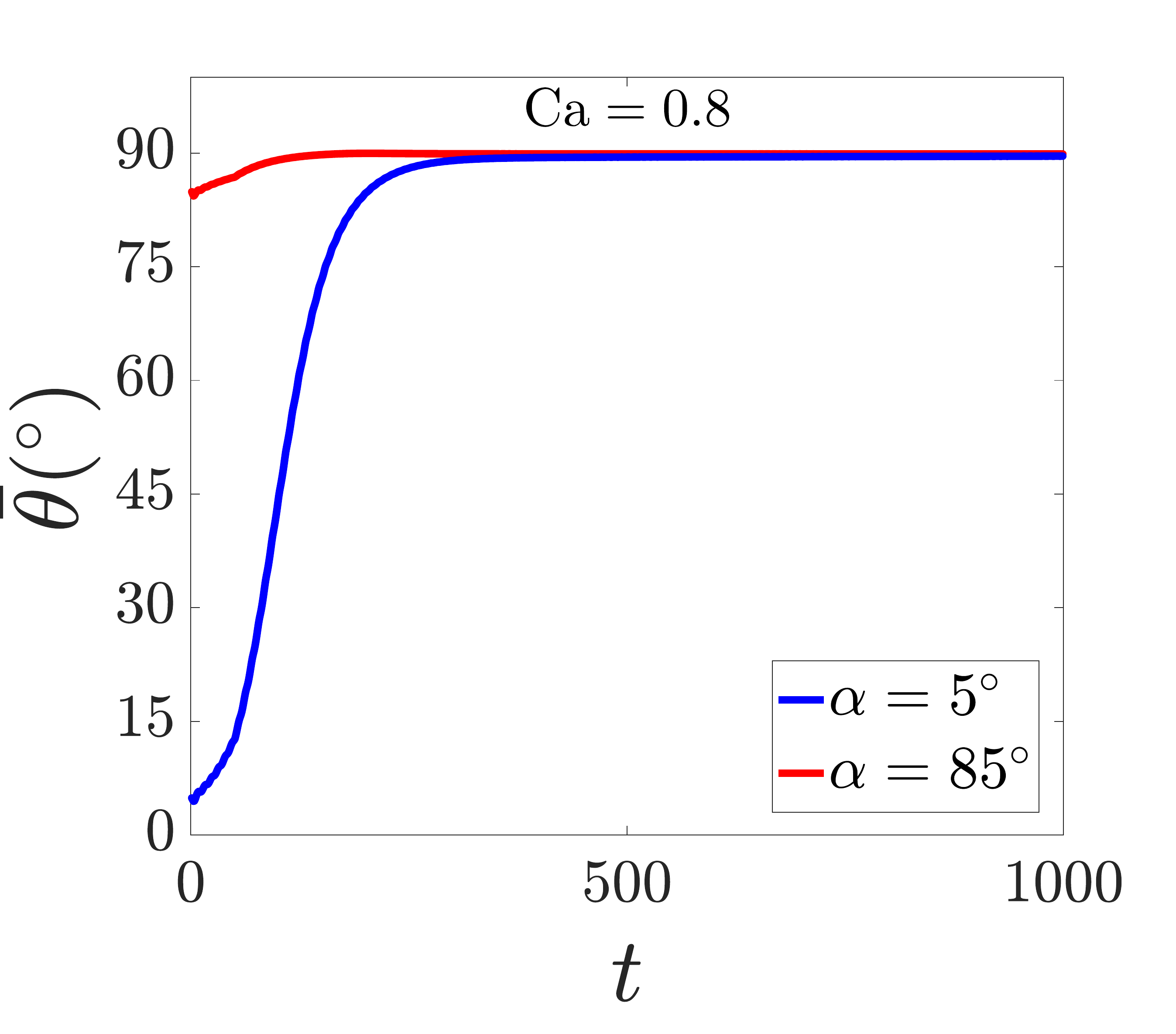}
    \label{fig:tumbling_to_swinging}
}
 \subfloat[]
{
    \includegraphics[width=0.4\textwidth]{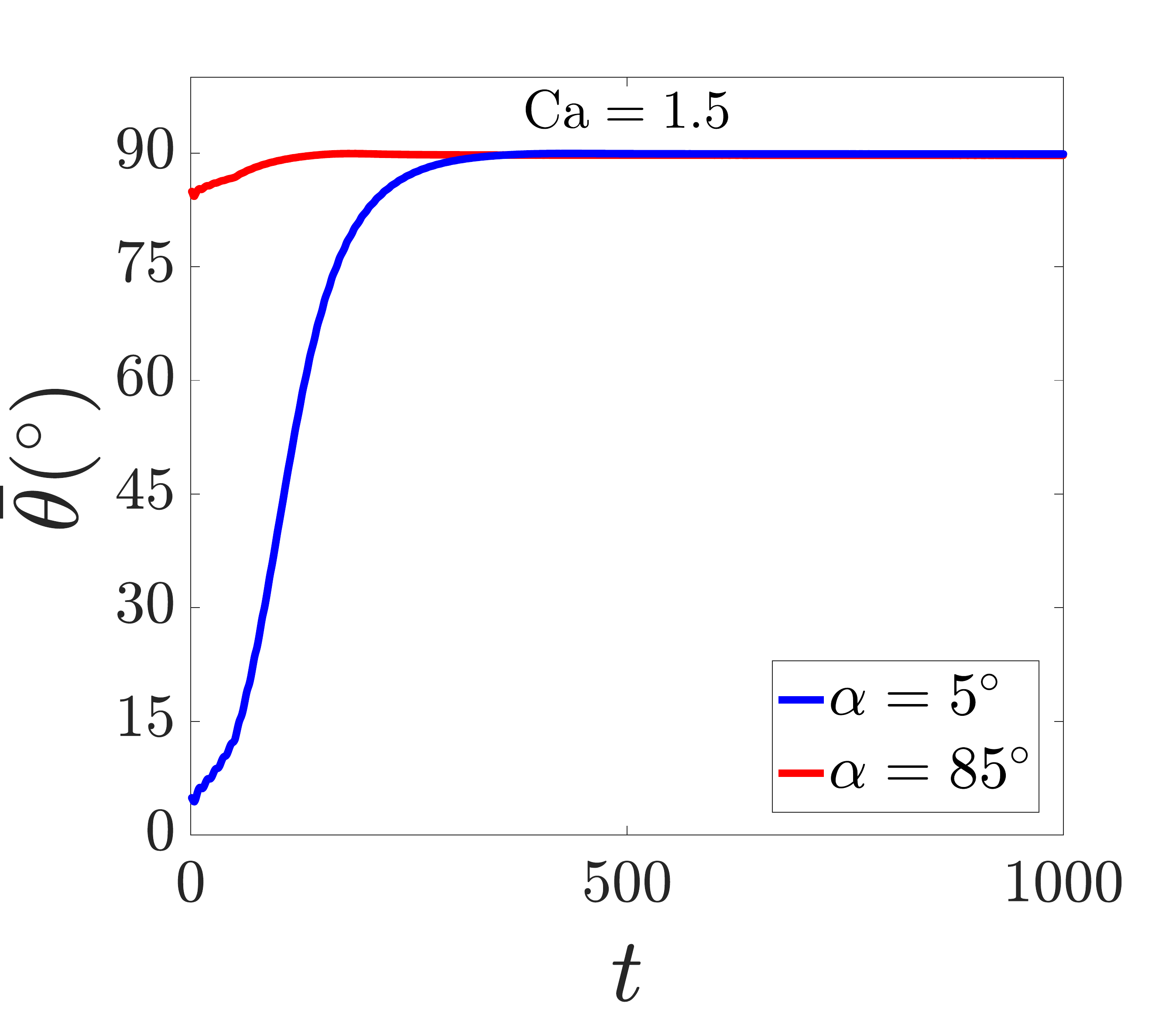}
    \label{fig:swinging}
}

\caption{Evolution of $\bar \theta$ of a prolate capsule (AR = 2.0) with $\hat{\kappa}_B = 0.2$ and $\lambda = 1$ at varying $\Ca$. At low $\Ca$ the capsule takes a stable tumbling motion at long times ((a) $\Ca$ = 0.08 and (b) $\Ca$ = 0.2)). A tumbling-to-swinging transition is observed at moderate $\Ca$ ((c) $\Ca$ = 0.8) before swinging becomes the attractor at high $\Ca$ ((d) $\Ca$ = 1.5).}
\label{lambda_1_large_bending}
\end{figure}

\begin{figure}[h]
\centering
\captionsetup{justification=raggedright}
 \subfloat[]
{
    \includegraphics[width=0.32\textwidth]{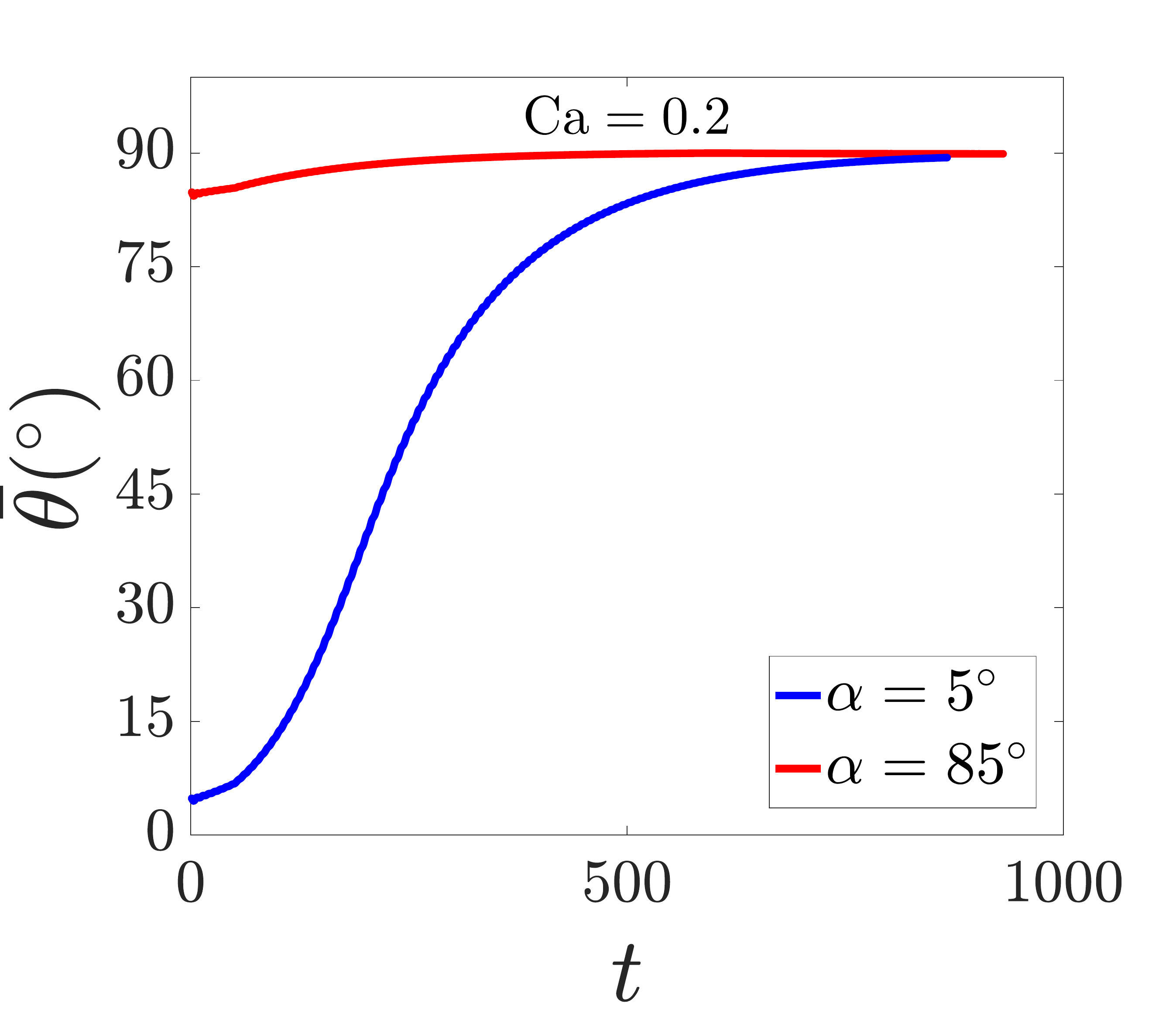}
    \label{fig:lambda_5_Ca_p2_large_bending}
}
 \subfloat[]
{
    \includegraphics[width=0.32\textwidth]{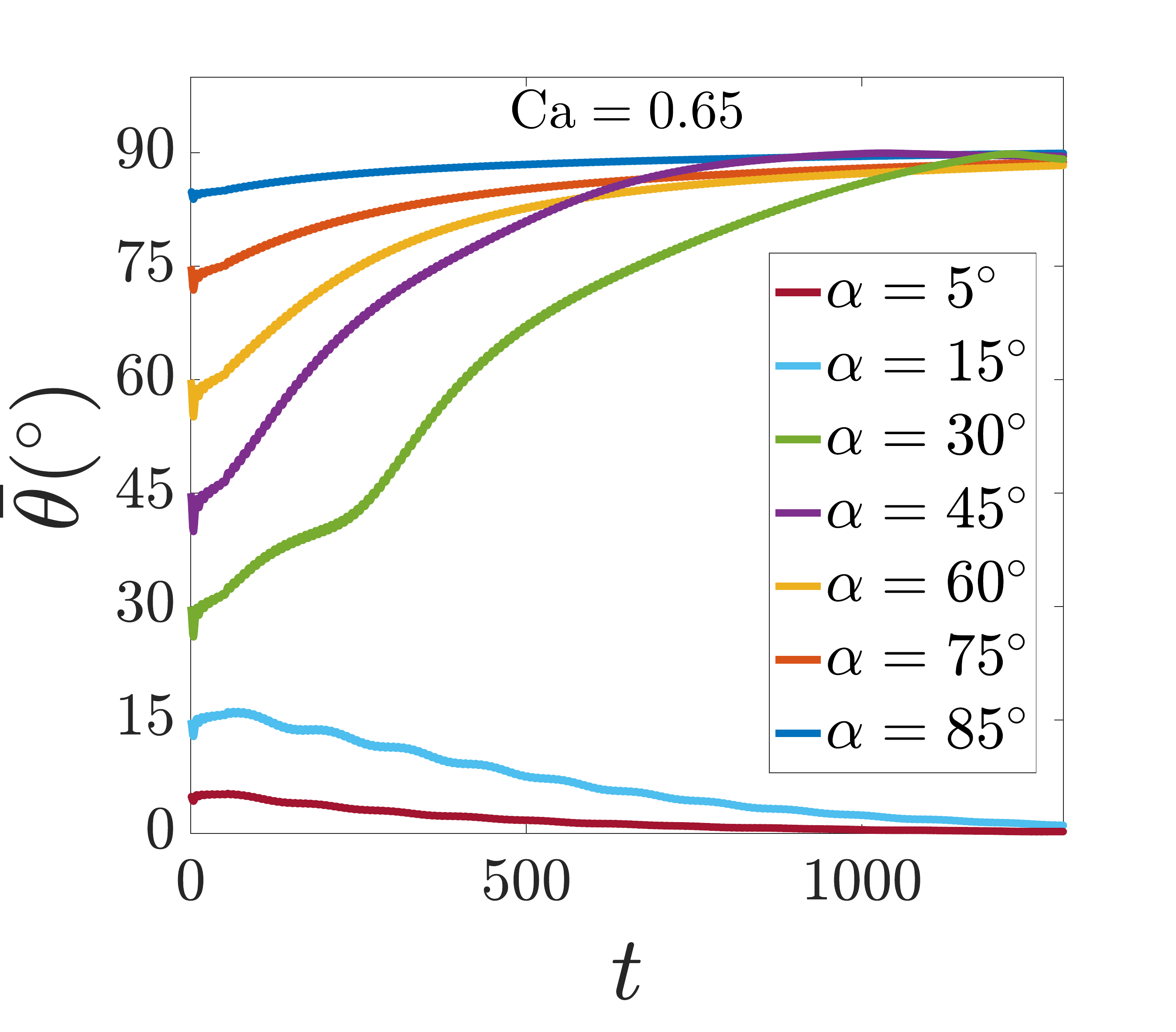}
    \label{fig:lambda_5_Ca_p65_large_bending}
}
 \subfloat[]
{
    \includegraphics[width=0.32\textwidth]{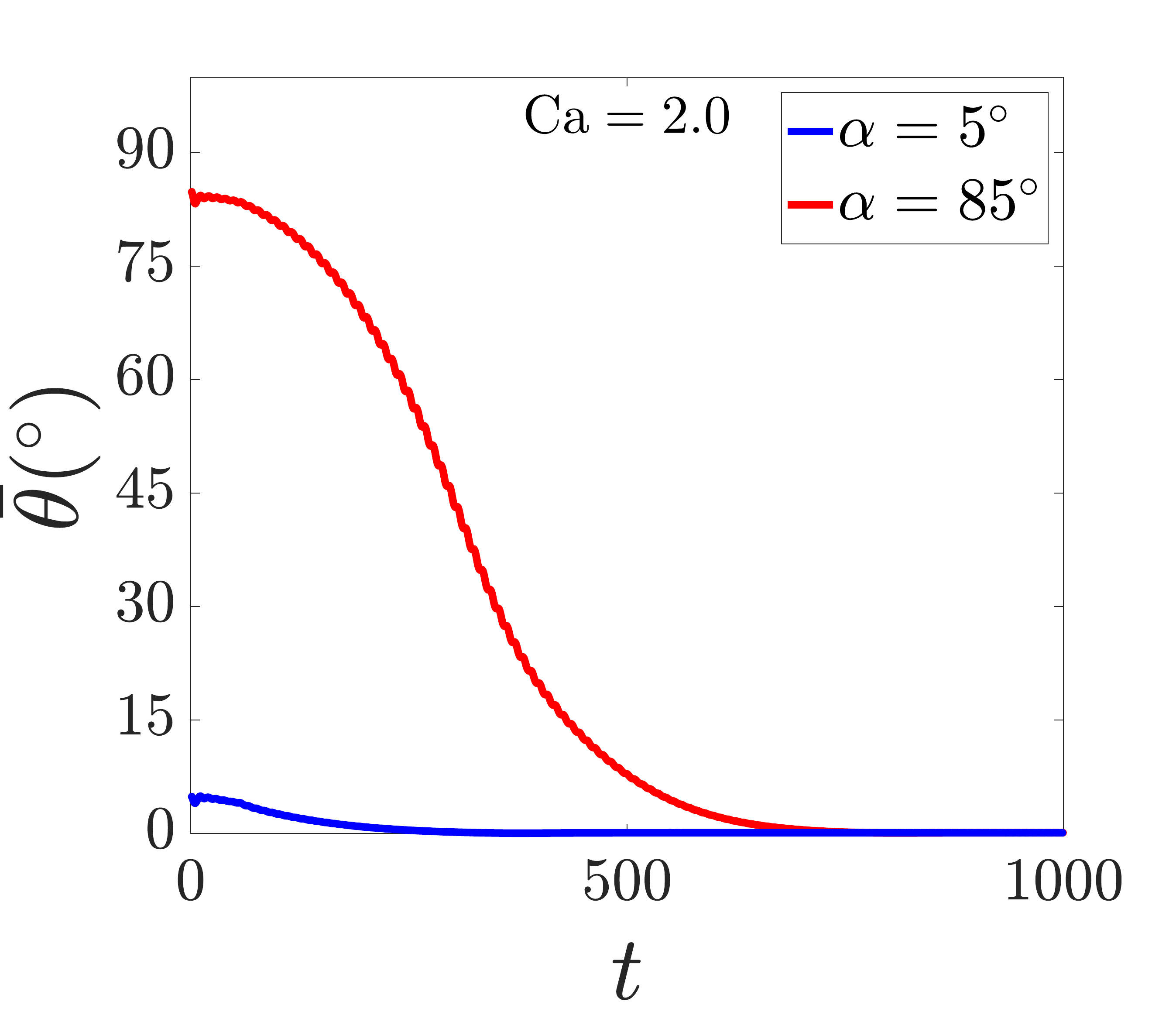}
    \label{fig:lambda_5_Ca_2_large_bending}
}

\caption{Evolution of $\bar \theta$ of a prolate capsule (AR = 2.0) with $\hat{\kappa}_B = 0.2$ and $\lambda = 5$ at (a) $\Ca$ = 0.2, (b) $\Ca$ = 0.65, and (c) $\Ca$ = 2.0.}
\label{lambda_5_large_bending}
\end{figure}

\begin{figure}[h]
\centering
\captionsetup{justification=raggedright}
 \subfloat[]
{
    \includegraphics[width=0.4\textwidth]{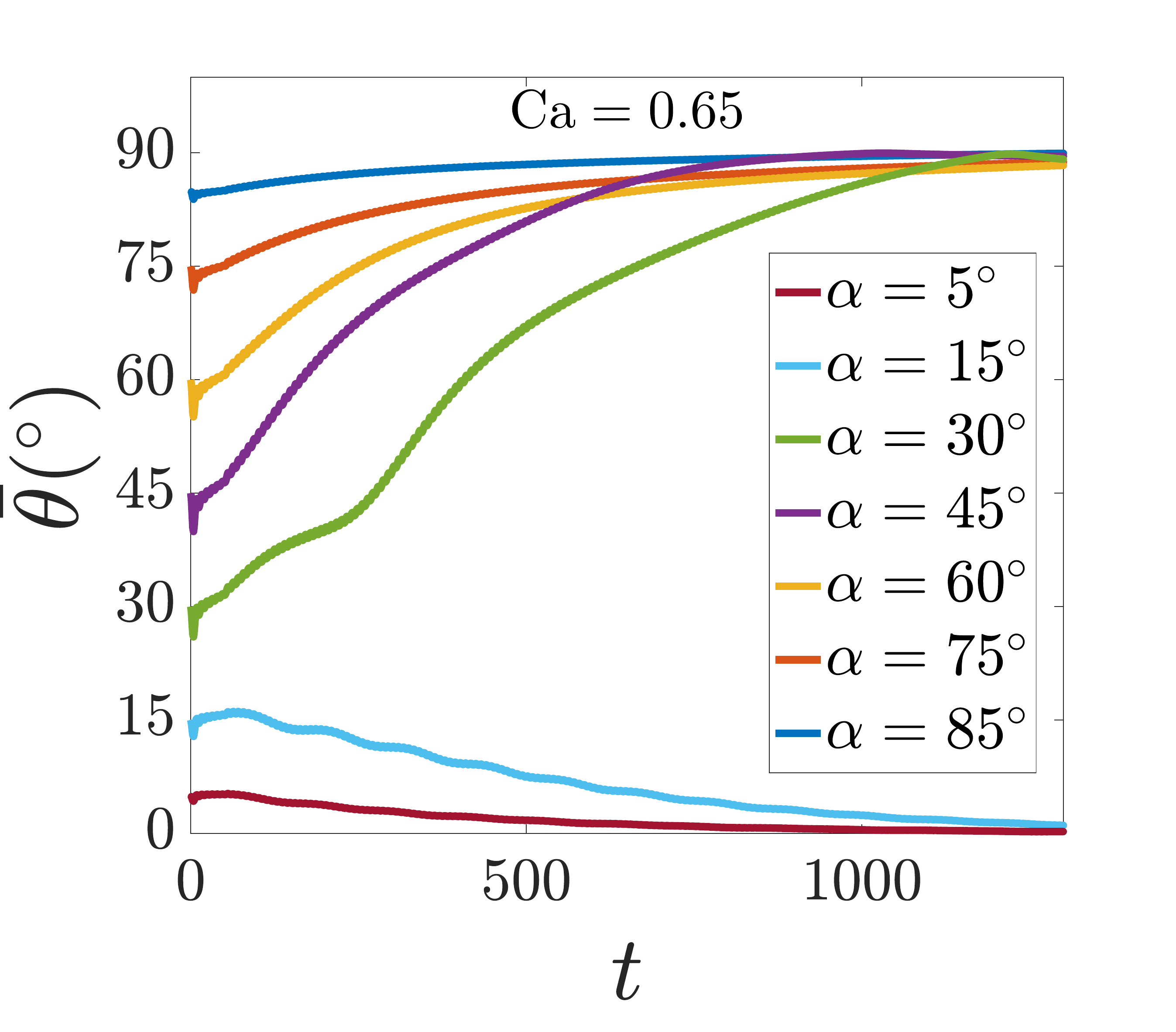}
    \label{fig:multiplicity_1}
}
 \subfloat[]
{
    \includegraphics[width=0.4\textwidth]{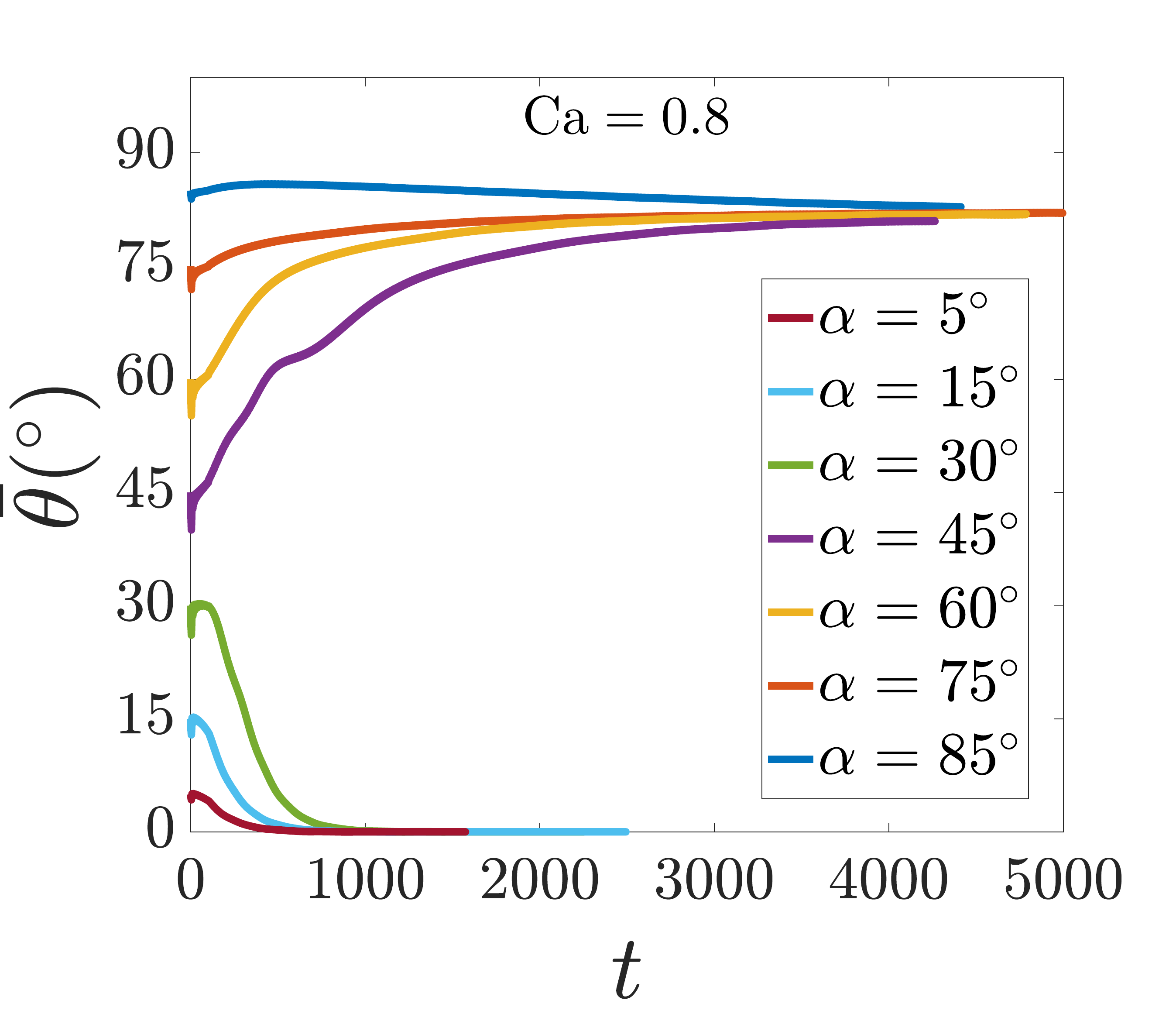}
    \label{fig:multiplicity_2}
}
\\
 \subfloat[]
{
    \includegraphics[width=0.4\textwidth]{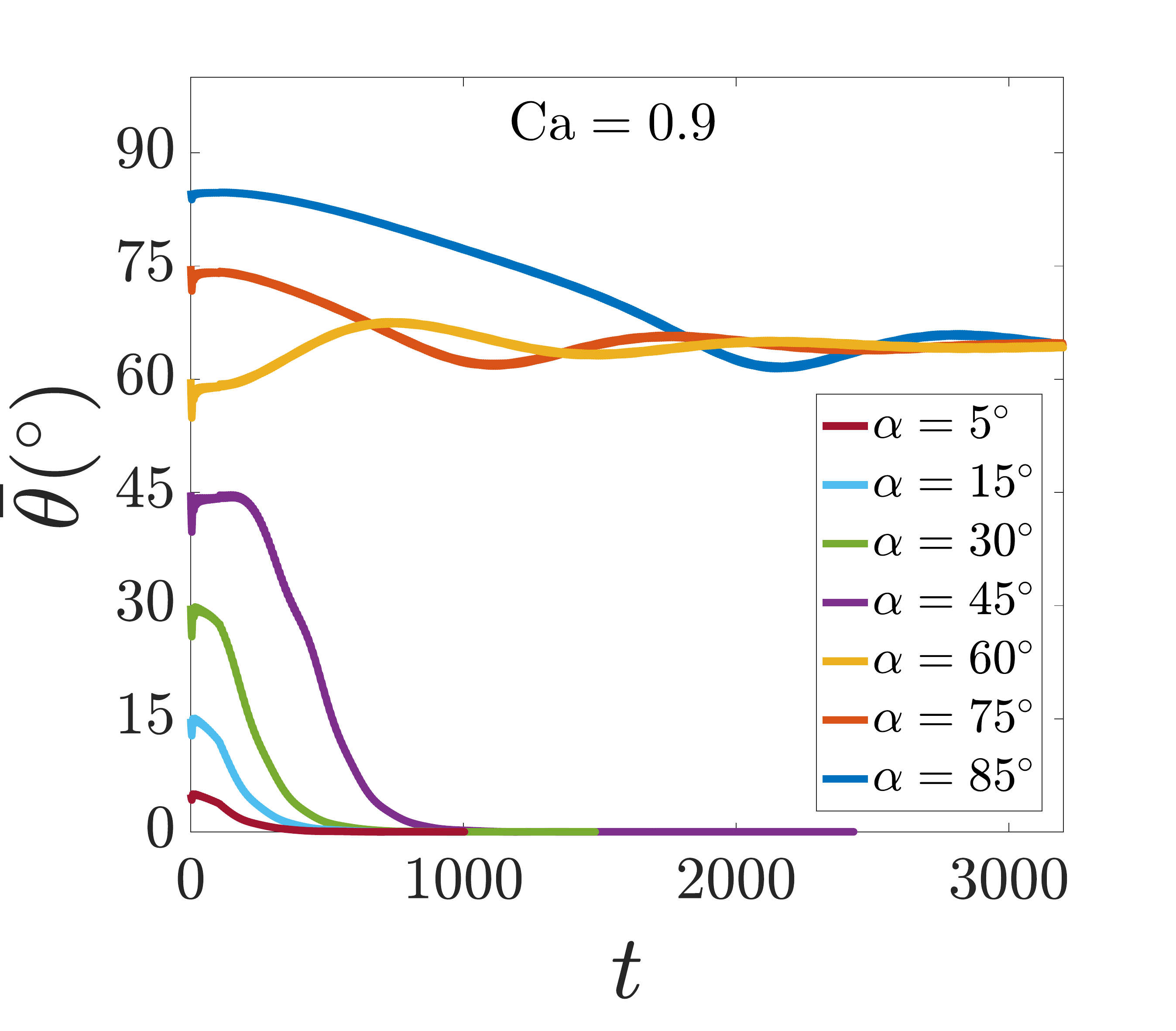}
    \label{fig:multiplicity_3}
}
 \subfloat[]
{
    \includegraphics[width=0.4\textwidth]{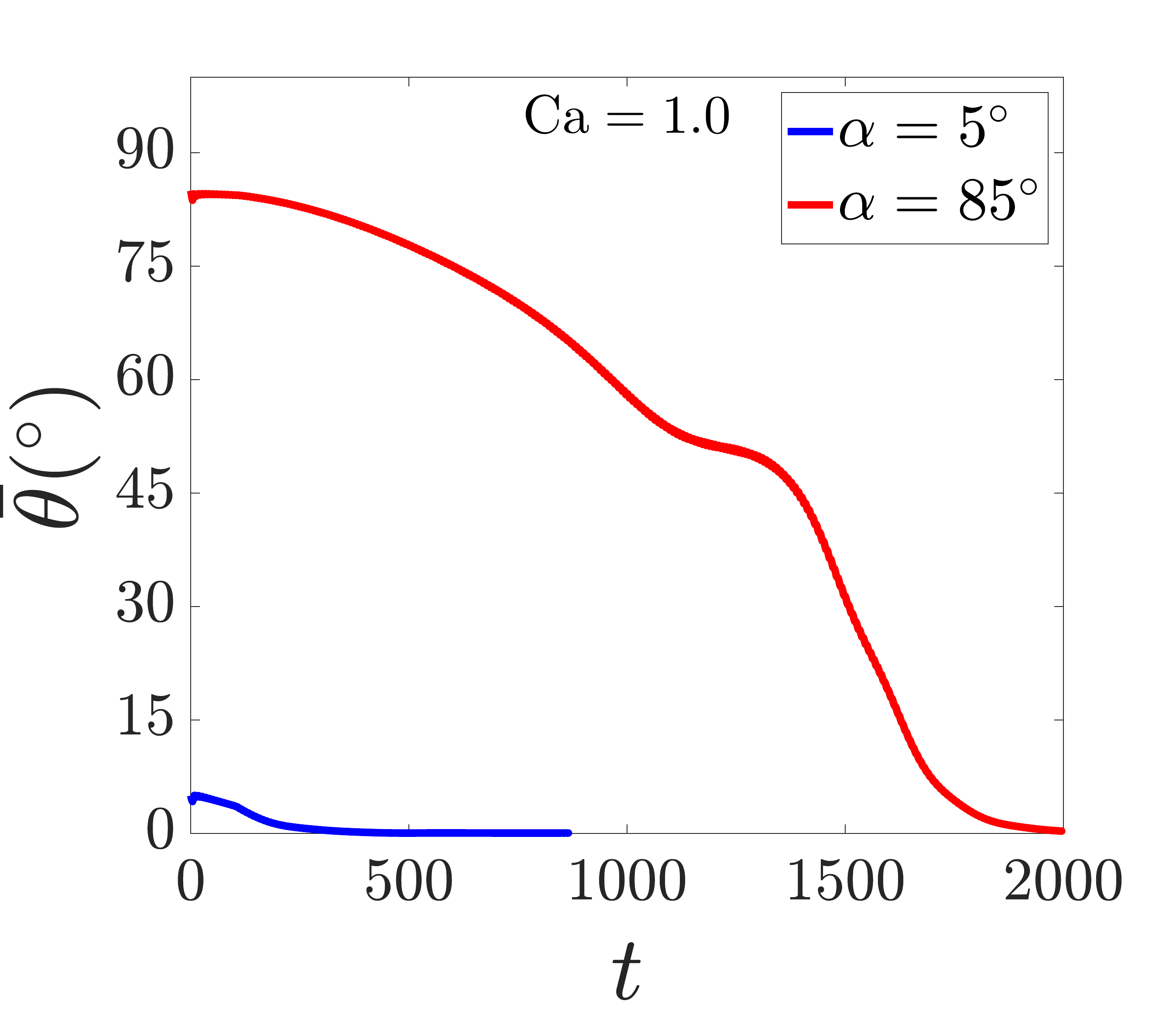}
    \label{fig:multiplicity_4}
}
\caption{Evolution of $\bar \theta$ of a prolate capsule (AR = 2.0) with $\hat{\kappa}_B = 0.2$ and $\lambda = 5$ at (a) $\Ca$ = 0.65, (b) $\Ca$ = 0.8, (c) $\Ca$ = 0.9, and (d) $\Ca$ = 1.0.}
\label{lambda_5_large_bending_bifurcation}
\end{figure}

\begin{figure}[h]
\centering
\captionsetup{justification=raggedright}
 \subfloat[]
{
    \includegraphics[width=0.46\textwidth]{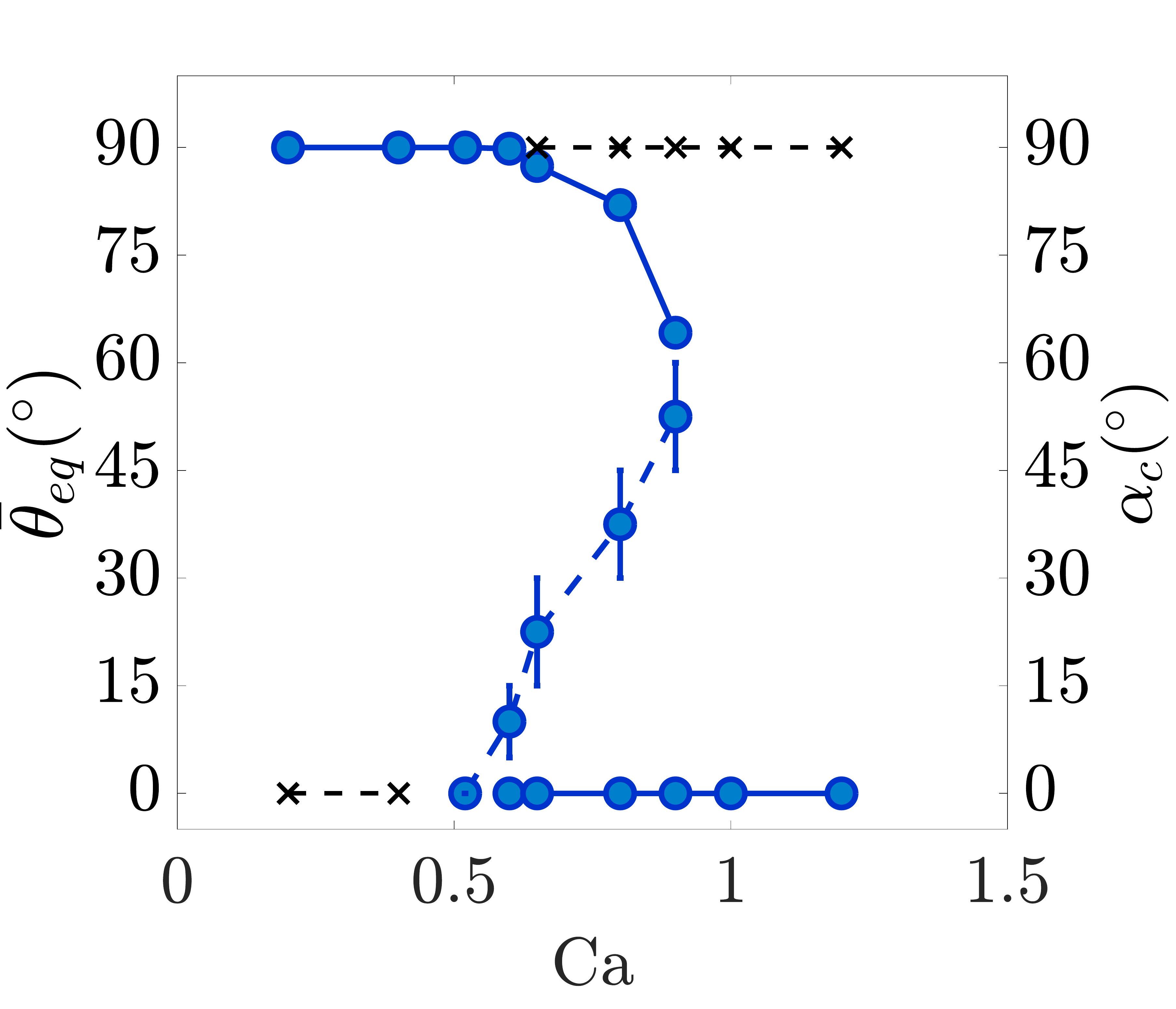}
    \label{fig:AR_2_lambda_5_bifurcation}
}
 \subfloat[]
{
    \includegraphics[width=0.49\textwidth]{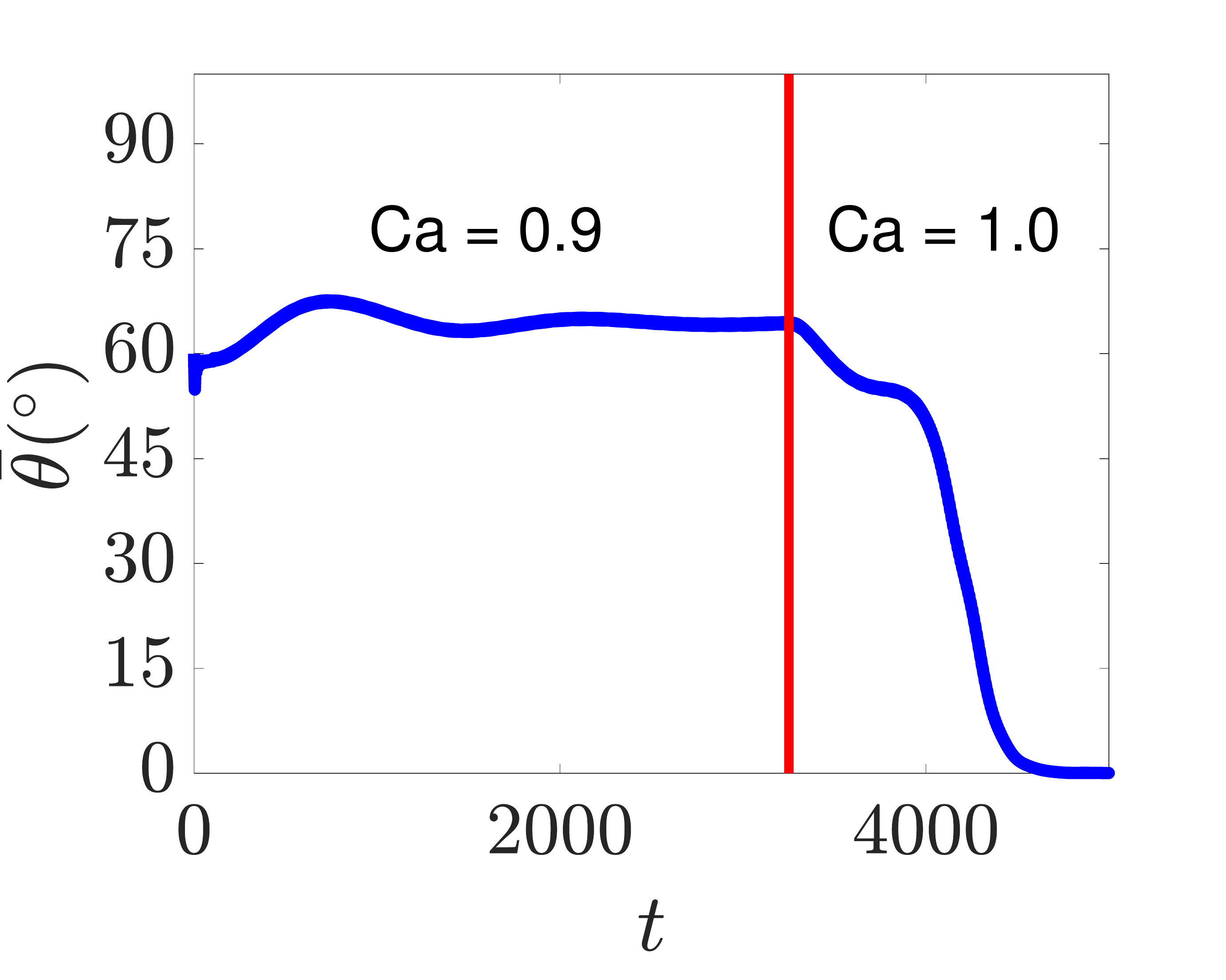}
    \label{fig:uniqueness_test}
}
\caption{(a) Bifurcation behavior for a prolate capsule (AR = 2.0) with $\hat{\kappa}_B = 0.2$ and $\lambda = 5$. The values for $\bar \theta_{eq}$ corresponding to the attractors are represented by the solid blue lines with filled circles, and the estimated values for $\alpha_c$ by the dashed blue line with filled circles (the error bars indicating the range for $\alpha_c$ at each $\Ca$ in the multiplicity regime). The unstable solutions at $\theta = 0^{\circ}$ and $\theta = 90^{\circ}$ are indicated by the dashed black lines with crosses. (b) Evolution of $\bar \theta$ for a capsule with $\alpha = 60^{\circ}$ initially at $\Ca = 0.9$. When a stable orbit is reached, the value of $\Ca$ is suddenly increased to 1.0, and the capsule evolves towards a new stable orbit.}
\label{uniqueness_test}
\end{figure}

\XZrevise{We found in the previous section that the long-time orbits for a prolate capsule with small bending stiffness are always independent of the initial conditions, although the stable orbital motions vary with capillary number.} In this section, we investigate the dynamics of a prolate capsule with large bending stiffness. The spontaneous shape of the capsule is always \XZrevise{assumed to be} the same as its rest shape unless stated otherwise. We vary the viscosity ratio $\lambda$, capillary number $\Ca$, initial orientation $\alpha$ and aspect ratio AR of the capsule while keeping the bending modulus $\hat{\kappa}_B = 0.2$. We reveal that large bending stiffness coupled with high viscosity ratio can give rise to a multiplicity \XZrevise{of} attractors for the capsule dynamics, which will be presented and discussed below. 

Again, we begin with $\lambda = 1$ and AR = 2.0. Unlike the case with $\hat{\kappa}_B = 0.02$\MDGrevise{,} in this case the stable orbit for the capsule is always in the shear plane, as shown by the $\alpha$-independent convergence of $\bar \theta$ towards $90^{\circ}$ at varying $\Ca$ in FIG.~\ref{lambda_1_large_bending}. The time it takes for this transient convergence decreases in general with increasing $\Ca$. This \XZrevise{conclusion} holds as $\lambda$ decreases from 1 to 0.2, and also as AR increases from 2.0 to 3.0 (not shown). In a previous numerical study by Walter \emph{et al.} \cite{walter-salsac-DBB-2011}, two regimes of stable orbital motion were found for a prolate capsule with the major axis initially lying in the shear plane: a rigid-body-like tumbling motion at low capillary number, and a fluid-like swinging motion at high shear rate\MDGrevise{,} in which the capsule elongation and orientation both oscillate in shear flow with the membrane continuously rotating around its deformed shape. Here we observe similar dynamics. At low $\Ca$, the capsule takes tumbling as the stable orbit (FIGs.~\ref{fig:tumbling_1} and \ref{fig:tumbling_2}), while at high $\Ca$ swinging becomes the attractor (FIG.~\ref{fig:swinging}); a tumbling-to-swinging transition is observed at moderate $\Ca$ (FIG.~\ref{fig:tumbling_to_swinging}), which is characterized by a nearly circular profile \XZrevise{of the deformed capsule} in the shear plane and roughly the same length of the two principal axes of the capsule during each half-period. \XZrevise{A similar transition was also observed for a prolate solid elastic particle in shear flow in both a theoretical prediction by Gao \emph{et al.} \cite{Gao2012} and a numerical investigation by Villone \emph{et al.} \cite{Villone2015}.}

Now we consider the cases where the viscosity ratio $\lambda$ is greater than unity. FIG.~\ref{lambda_5_large_bending} shows the evolution of $\bar \theta$ over time for a prolate capsule with AR = 2.0 and $\lambda = 5$ at varying $\Ca$. At low and high $\Ca$ regimes, the orbit of the capsule converges towards $\bar \theta_{eq} = 90^{\circ}$ and $\bar \theta_{eq} = 0^{\circ}$, representing a stable tumbling and rolling motion, respectively (FIGs.~\ref{fig:lambda_5_Ca_p2_large_bending} and \ref{fig:lambda_5_Ca_2_large_bending}). At intermediate $\Ca$, however, the orbit taken by the capsule at long times is found to depend on the initial orientation, as observed in FIG.~\ref{fig:lambda_5_Ca_p65_large_bending}. \XZ{The reason I didn't refer to the bifurcation diagram here is that I am now describing FIG.~\ref{lambda_5_large_bending_bifurcation} first. The concepts of bifurcation will be introduced in the next paragraph.}Several values of $\alpha$ ranging from $5^{\circ}$ to $85^{\circ}$ are considered here to illustrate the dependence of the long-time orbit on the initial orientation, and the results for a range of $\Ca$ are displayed in FIG.~\ref{lambda_5_large_bending_bifurcation}. The key characteristic of this regime is \XZrevise{that multiple stable orbits coexist, as observed in FIGs.~\ref{fig:multiplicity_1}, \ref{fig:multiplicity_2} and \ref{fig:multiplicity_3}. In this multiplicity regime, two stable solution branches are found as the capsule with varying $\alpha$ evolves towards the long-time orbit.} The lower branch always corresponds to a stable rolling motion with $\bar \theta_{eq} = 0^{\circ}$. For the upper branch, $\bar \theta_{eq}$ decreases with increasing $\Ca$ from $90^{\circ}$, which corresponds to a stable tumbling motion, to intermediate values representing stable precessing motions about the shear plane. The multiplicity \XZrevise{of} attractors disappears as $\Ca$ further increases, and $\bar \theta_{eq} = 0^{\circ}$ (rolling) becomes the only stable orbit (FIG.~\ref{fig:multiplicity_4}). Another important observation is that the critical value of $\alpha$ for the two branches of the evolution results, denoted as $\alpha_c$, increases as $\Ca$ increases. Here it is not necessary to determine the exact values for $\alpha_c$; actually, a range for $\alpha_c$ at each $\Ca$ is sufficient to show this trend: $\alpha_c \in (15^{\circ}, 30^{\circ})$ at $\Ca$ = 0.65 (FIG.~\ref{fig:multiplicity_1}), $\alpha_c \in (30^{\circ}, 45^{\circ})$ at $\Ca$ = 0.8 (FIG.~\ref{fig:multiplicity_2}), and $\alpha_c \in (45^{\circ}, 60^{\circ})$ at $\Ca$ = 0.9 (FIG.~\ref{fig:multiplicity_3}).

\XZ{I think it's better to start this part in a new paragraph - it introduces the idea of bifurcations. Also the previous paragraph would be too long if they were combined.}The global effects of $\Ca$ on $\bar \theta_{eq}$ for the attractors and $\alpha_c$ for the multiplicity regime described above are summarized \MDGrevise{as a bifurcation diagram} in FIG.~\ref{fig:AR_2_lambda_5_bifurcation}. The values for $\bar \theta_{eq}$ corresponding to the attractors in different regimes, either single or multiple, are represented by the solid blue lines with circles, and the estimated values for $\alpha_c$ by the dashed blue line with circles. \XZrevise{Again, there are always solutions at $\theta = 0^{\circ}$ and $\theta = 90^{\circ}$ that correspond to the orbits for a capsule aligned with the $z$ axis and in the shear plane, respectively. These solutions are found to be unstable in certain regimes of $\Ca$, as indicated by the dashed black lines with crosses.} \XZrevise{Indeed, there is a symmetry-breaking bifurcation away from $\theta = 90^{\circ}$ between $\Ca = 0.6$ and $\Ca = 0.65$, and away from $\theta = 0^{\circ}$ between $\Ca = 0.4$ and $\Ca = 0.52$. Two branches of attractors are observed.} With any instantaneous orientation $\theta$ above the dashed blue line, the capsule would be attracted to a stable orbit \XZrevise{on the upper branch of the attractors} at the corresponding $\Ca$, while any value for $\theta$ below the dashed blue line converges towards a rolling orbit on the lower branch with $\bar \theta_{eq}=0^\circ$. \MDGrevise{The turning point at $\Ca \approx 0.9$ corresponds to a saddle-node bifurcation -- for higher $\Ca$ this intermediate solution loses existence.} \MDGrevise{This loss of existence is indicated by computing the solution at $\Ca$ = 0.9 and suddenly increasing} $\Ca$ from 0.9 to 1.0 \MDGrevise{(FIG.~\ref{fig:uniqueness_test}).} \MDGrevise{The capsule now evolves} from a stable precession to a rolling motion as taken by a capsule with the same initial orientation initially at $\Ca$ = 1.0. \XZrevise{Similar to our findings, a tumbling-to-rolling transition was also observed for a prolate capsule (AR = 2.0, $\lambda = 1$) with zero bending resistance but in the presence of a small particle inertia \cite{Wang2013}, and a multiplicity of attractors was found in the transition regime.} Other studies \cite{Yu2007,huang_yang_krafczyk_lu_2012,rosen_lundell_aidun_2014} have also reported a dependence of the stable orbit on the initial orientation for a solid prolate particle at certain Reynolds number regimes. \XZ{I think this paragraph should be combined with the next.} FIG.~\ref{fig:attractor_and_critical_alpha_various_AR} shows the \XZrevise{bifurcation behavior of the attractors} for a prolate capsule ($\lambda = 5$) with various aspect ratios. It is observed that as AR increases from 2.0 to 3.0, the onset of the multiplicity regime occurs at a higher $\Ca$, and this regime becomes broader. The effect of the viscosity ratio $\lambda$ is illustrated in FIG.~\ref{fig:attractor_and_critical_alpha_AR_3_varying_lambda} \XZrevise{for a capsule with AR = 3.0}. In the parameter regime considered here, the multiplicity regime starts at a lower $\Ca$ and becomes narrower with increasing $\lambda$, showing an effect opposite to that of the aspect ratio. The orbital modes \XZrevise{are determined for the attractors on the upper branch of the multiplicity, and} summarized in a phase diagram over a wide range of \XZrevise{$\Ca$ and} $\lambda$ (FIG.~\ref{fig:phase_diagram_lambda_over_Ca}). \XZrevise{Snapshots of a capsule ($\lambda = 6$) taking a stable tumbling, precessing, and rolling motion are shown in FIGs.~\ref{fig:tumbling_snapshots}, \ref{fig:precessing_snapshots} and \ref{fig:rolling_snapshots}, respectively, as examples.} 

  \begin{figure}[h]
  \centering
  \captionsetup{justification=raggedright}
  \subfloat[]
  {
    \includegraphics[width=0.46\textwidth]{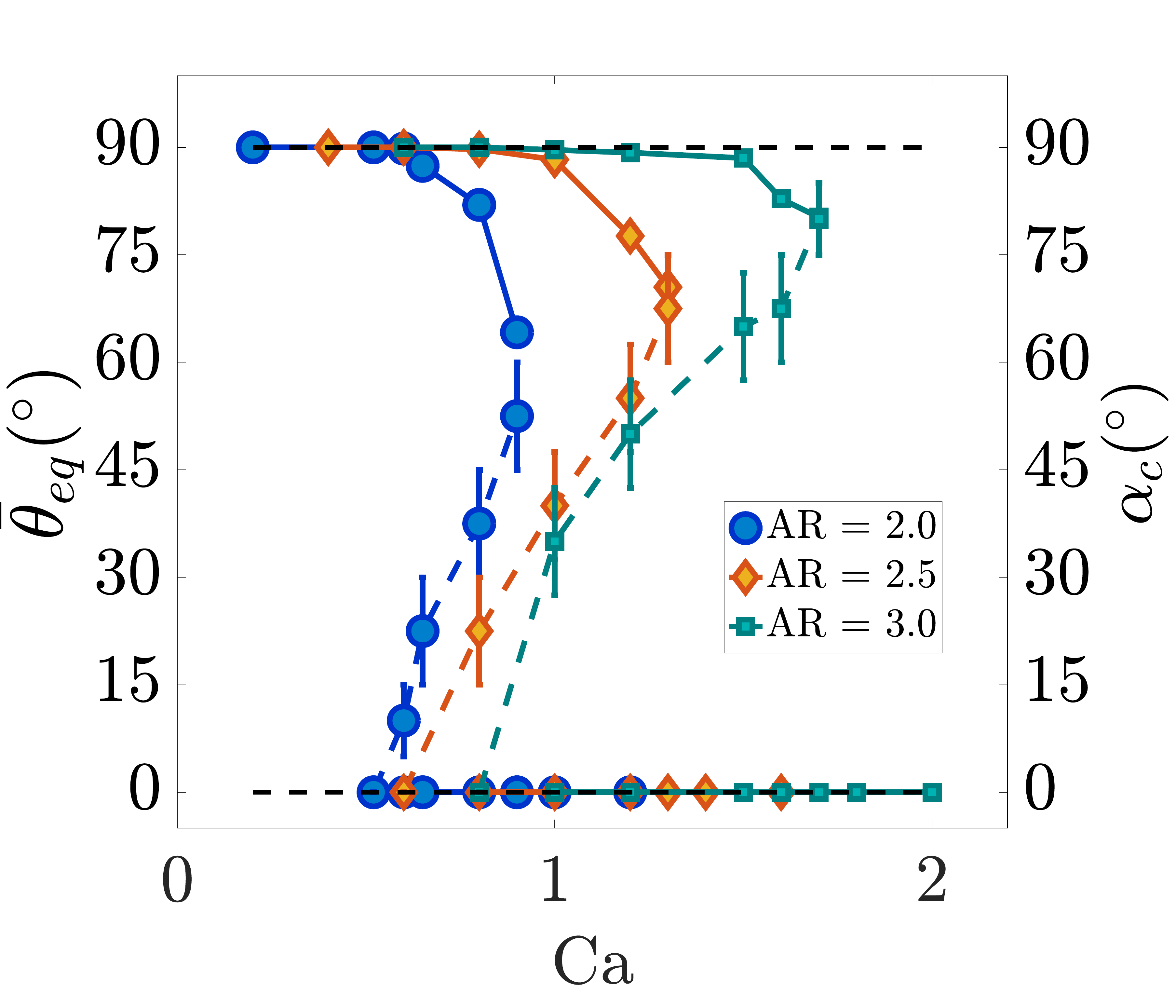}
    \label{fig:attractor_and_critical_alpha_various_AR}
  }
   \subfloat[]
  {
    \includegraphics[width=0.46\textwidth]{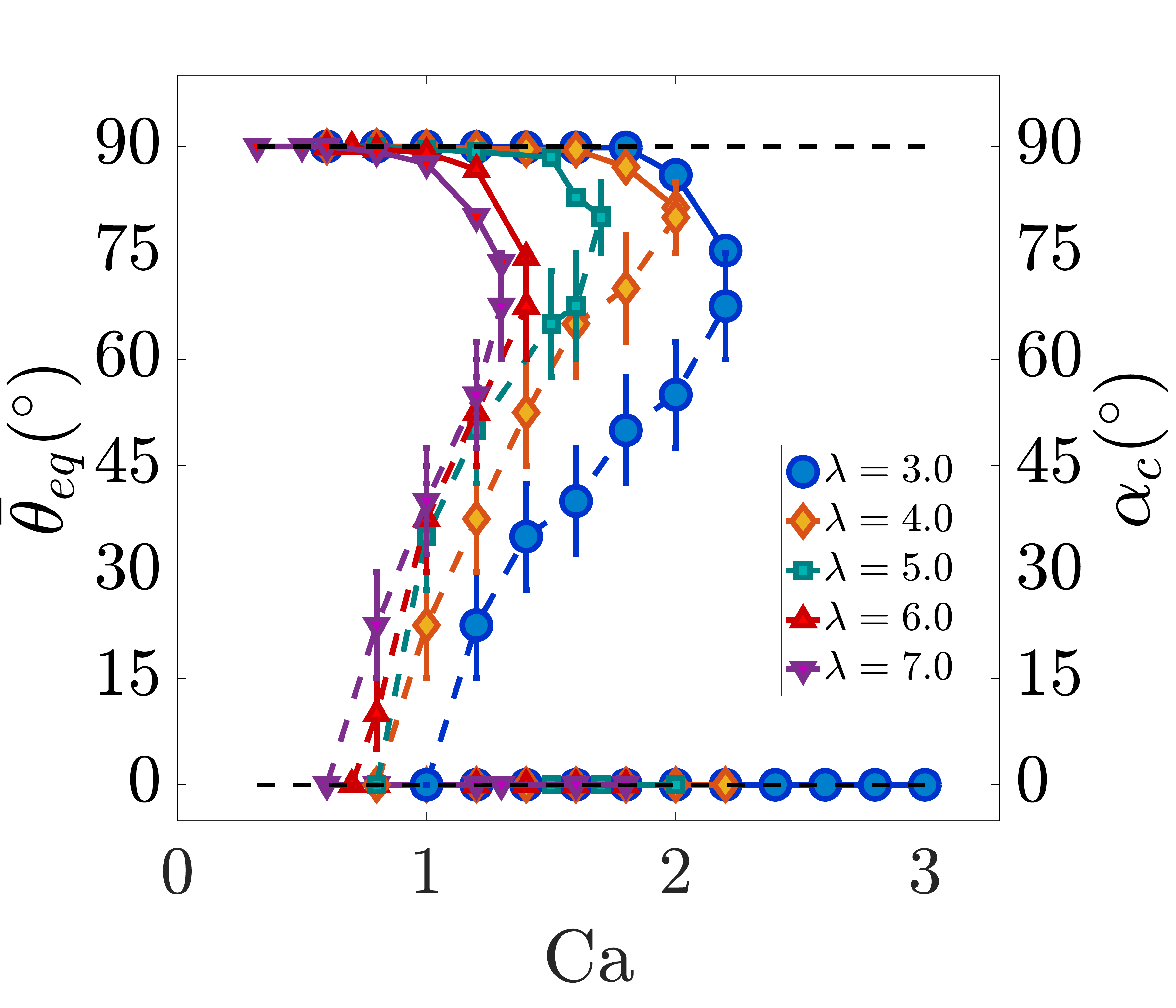}
    \label{fig:attractor_and_critical_alpha_AR_3_varying_lambda} 
  }
  \\
   \subfloat[]
{
    \includegraphics[width=0.45\textwidth]{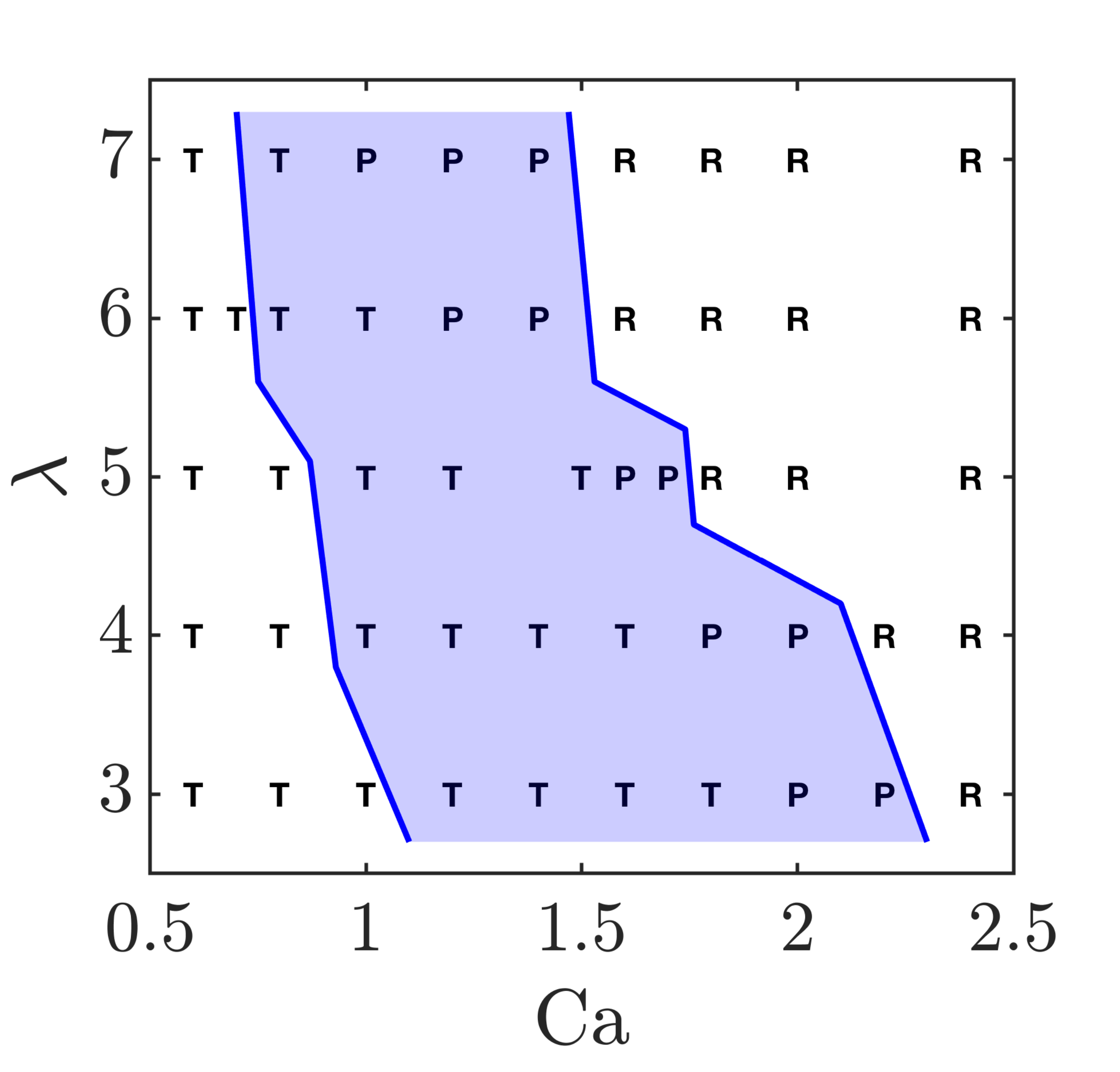}
    \label{fig:phase_diagram_lambda_over_Ca}
}

  \caption{Effects of capsule aspect ratio (a, $\lambda = 5$) and the viscosity ratio (b, AR = 3.0) on the bifurcation behavior of the attractors for a prolate capsule with $\hat{\kappa}_B = 0.2$. Again, the black dashed lines at $\theta = 0^{\circ}$ and $\theta = 90^{\circ}$ correspond to the orbits for a capsule aligned with the $z$ axis and in the shear plane, respectively. (c) Phase diagram of stable orbital motions for a prolate capsule (AR = 3.0) with $\hat{\kappa}_B = 0.2$ over a range of $\Ca$ and $\lambda$. The multiplicity regime is shaded in blue, and only the motion modes for the attractors on the upper branch are shown here (the attractors on the lower branch always correspond to rolling and thus are not shown here). T denotes tumbling, P precessing, and R rolling.}
  \label{alpha_theta_bifurcation}
  \end{figure}

\begin{figure}[h]
\centering
\captionsetup{justification=raggedright}
 \subfloat[]
{
    \includegraphics[width=0.75\textwidth]{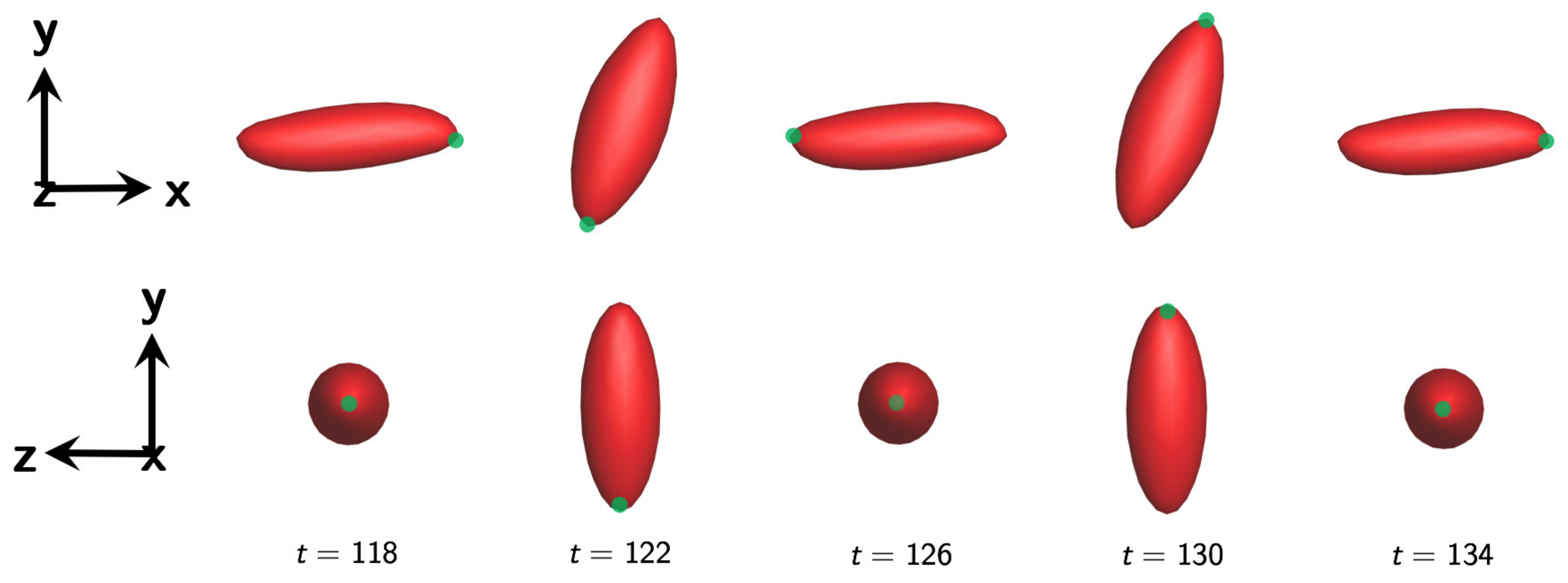}
    \label{fig:tumbling_snapshots}
} 
\\
\subfloat[]
{
    \includegraphics[width=0.75\textwidth]{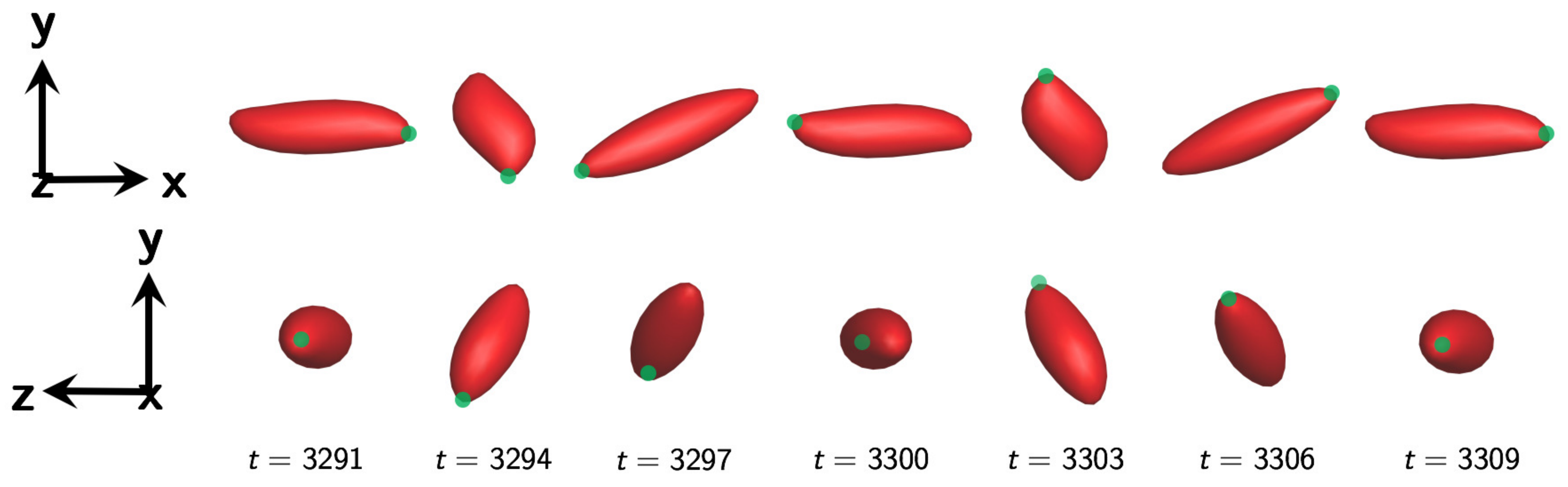}
    \label{fig:precessing_snapshots}
}
\\
\subfloat[]
{
    \includegraphics[width=0.75\textwidth]{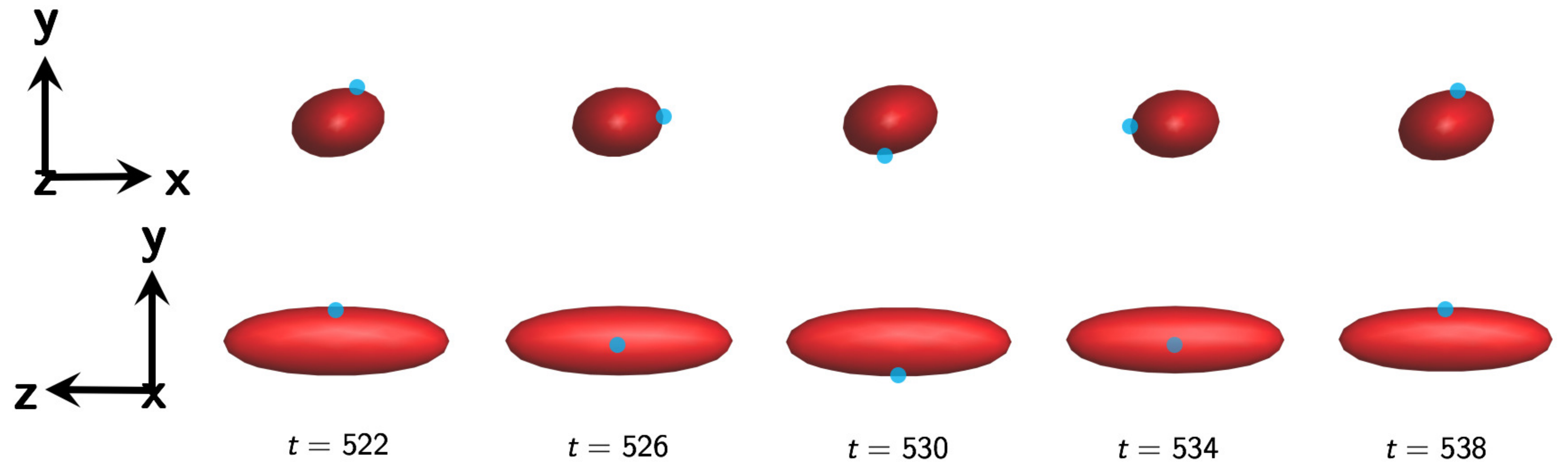}
    \label{fig:rolling_snapshots}
}
\caption{Time sequence images (side and front views) of a prolate capsule (AR = 3.0) with $\hat{\kappa}_B = 0.2$ and $\lambda = 6.0$ taking a tumbling (a, $\Ca$ = 0.6), precessing (b, $\Ca$ = 1.4), and rolling (c, $\Ca$ = 2.0) motion, respectively. \XZrevise{The initial orientation of the capsule is $\alpha = 75^{\circ}$. The markers indicate the positions of a membrane point initially on the major axis of the capsule for (a) and (b), and a membrane point initially on the equator of the capsule for (c).}} 
\label{snapshots_high_bending}
\end{figure}

The multiplicity in attractors for a single prolate capsule, as described above, implies a multiplicity in the rheological properties for a dilute suspension of such capsules (AR = 3.0 and $\hat{\kappa}_B = 0.2$). Results are summarized in FIG.~\ref{rheology_lambda_5} for two viscosity ratios, $\lambda = 3.0$ and $\lambda = 5.0$. For each $\lambda$, the squares represent the pure tumbling regime at lower $\Ca$, while the circles represent the pure rolling regime at higher $\Ca$; the multiplicity regime at intermediate $\Ca$ is bounded by vertical dotted lines, the downward and upward triangles representing the attractors on the upper branch (either tumbling or precessing) and the stable rolling orbits on the lower branch, respectively. All capsules are assumed to take the same stable orbit at each $\Ca$. \MDGrevise{In this idealized situation, two different stable values for the particle stress can coexist when $\Ca$ is in the multiplicity regime.} A general shear-thinning behavior is observed in FIG.~\ref{fig:Sigma_xy} for both viscosity ratios, and minor differences are observed between these two cases. \XZrevise{This shear-thinning behavior is also observed for $\lambda = 1.0$ and $\lambda = 0.2$ (not shown).} Another observation is that $[\eta]$ is higher when the capsules are taking a stable tumbling (or precessing) motion than it is when they are rolling. Mueller \emph{et al.} \cite{Mueller2010}\sout{extended Jeffery's work \cite{Jeffery:1922wb} and} numerically determined the Einstein coefficient for a dilute suspension of inertialess non-spherical solid particles with a distribution of orientations, and reported similar findings: the contribution of a prolate particle to the suspension viscosity (i.e. the $yx$-component of the particle's force dipole) decreases as \sout{the orbit constant decreases from $C = \infty$ (tumbling) to $C = 0$ (rolling)}\XZrevise{the orbit of the capsule evolves from tumbling to rolling}. Huang \emph{et al.} \cite{huang_yang_krafczyk_lu_2012} showed that this conclusion still holds in the presence of small inertia. For the dimensionless normal stress differences $N_1$ and $N_2$ (FIGs.~\ref{fig:N1} and \ref{fig:N2}), again, the values are always positive for $N_1$ and negative for $N_2$, with $|N_2|\ll N_1$. Specifically, $N_1$ displays an obvious non-monotonic dependency on $\Ca$ when the capsules are tumbling or precessing, while having smaller values and showing a small variation with $\Ca$ for a suspension of rolling capsules. This trend with $\Ca$ is qualitatively similar for both viscosity ratios, even though the magnitude of $N_1$ corresponding to each line is much smaller for the higher viscosity ratio ($\lambda = 5.0$). A non-monotonicity is also observed for $N_2$. \XZrevise{Overall, as a result of the existence of the multiplicity regime, the values for each quantity of the rheology follow the path along the upper branch, and eventually fall onto the lower branch through the right vertical dotted line upon quasi-statically increasing $\Ca$; conversely, the values move along the lower branch and jump onto the upper branch via the left vertical dotted line upon quasi-statically decreasing $\Ca$. These processes are illustrated by the arrows in each figure of FIG.~\ref{rheology_lambda_5}.} 

\begin{figure}[h]
\centering
\captionsetup{justification=raggedright}
 \subfloat[]
{
    \includegraphics[width=0.45\textwidth]{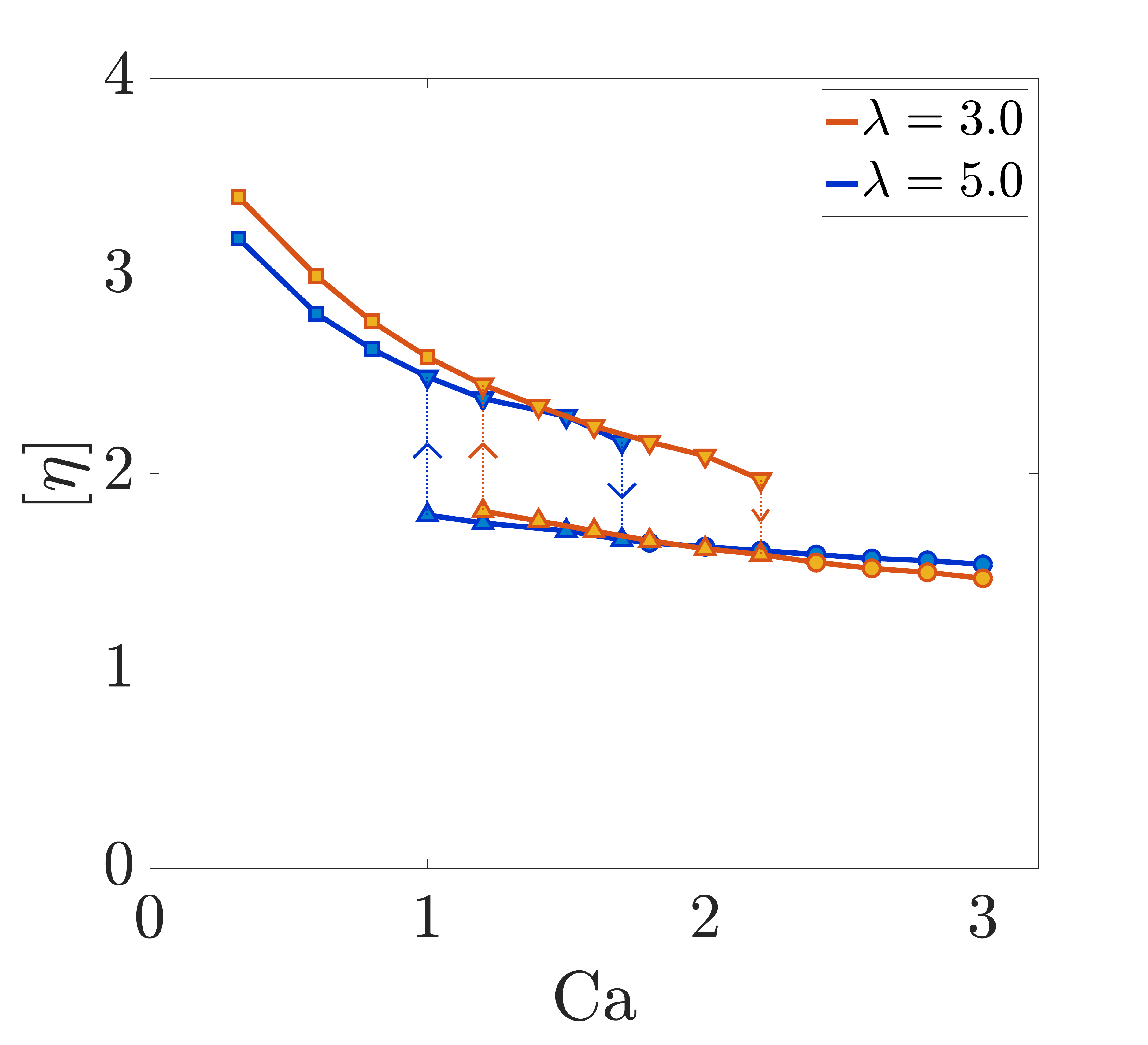}
    \label{fig:Sigma_xy}
}
\\
 \subfloat[]
{
    \includegraphics[width=0.45\textwidth]{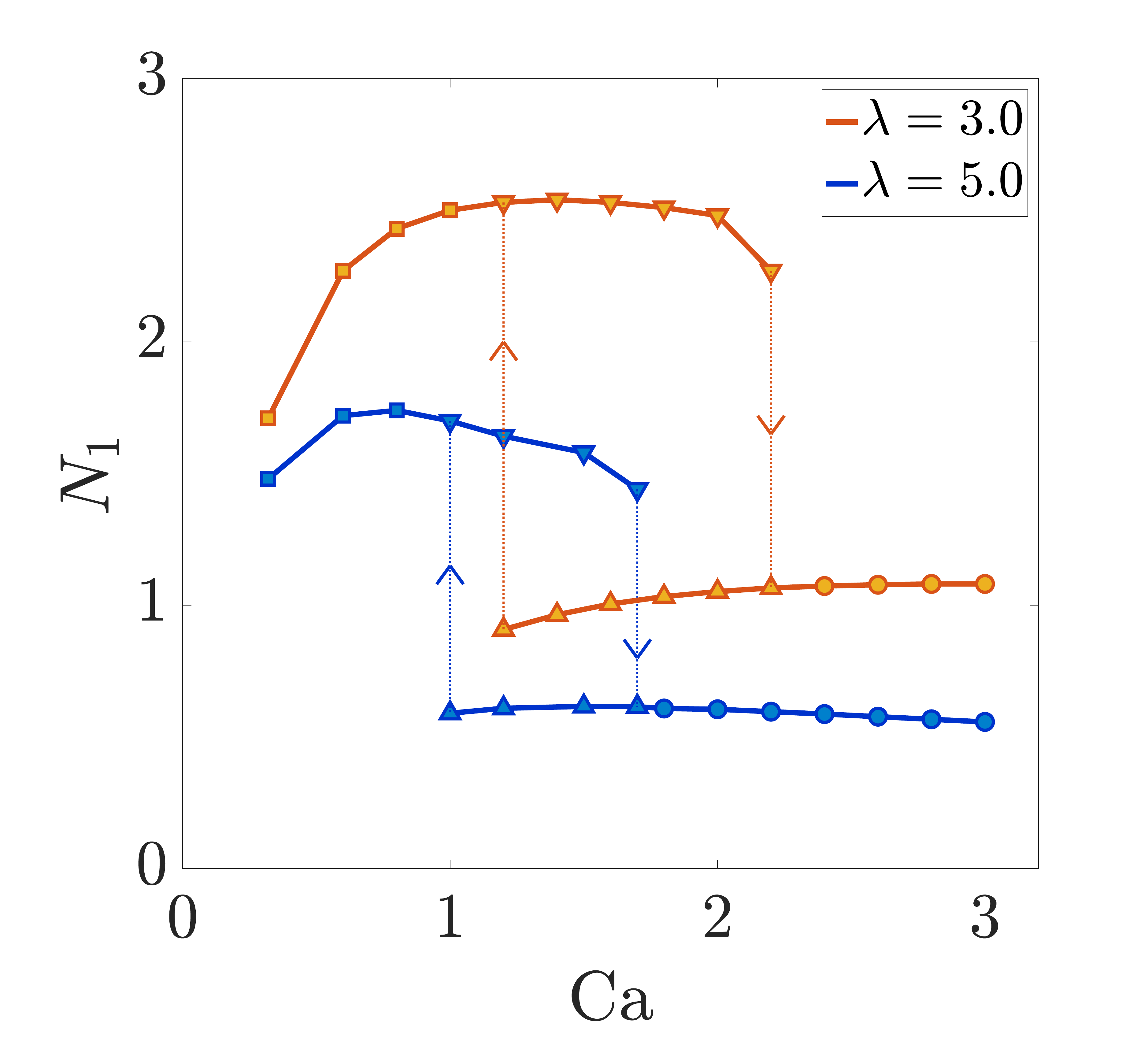}
    \label{fig:N1}
}
 \subfloat[]
{
    \includegraphics[width=0.45\textwidth]{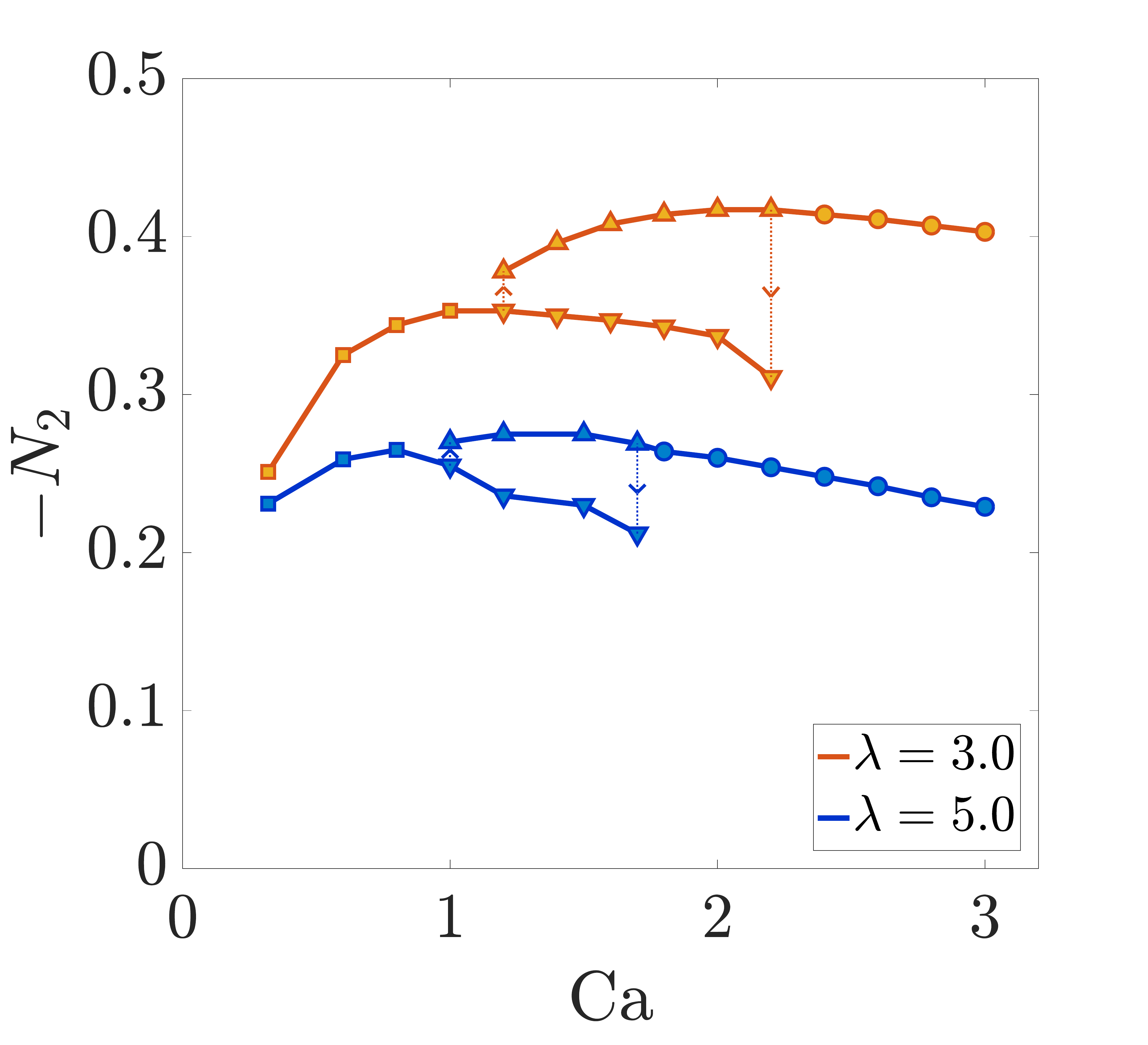}
    \label{fig:N2}
}

\caption{Intrinsic viscosity (a) and dimensionless normal stress differences (b,c) predicted for a dilute suspension of identical prolate capsules (AR = 3.0) with $\hat{\kappa}_B = 0.2$ taking the same corresponding stable orbit(s) at varying $\Ca$ for $\lambda = 3.0$ and $\lambda = 5.0$. In each case, the squares represent the pure tumbling orbits at lower $\Ca$, and the circles represent the pure rolling orbits at higher $\Ca$; the downward and upward triangles represent the attractors on the upper branch (either tumbling or precessing) and the stable rolling orbits on the lower branch of the multiplicity regime (bounded by vertical dotted lines), respectively.}
\label{rheology_lambda_5}
\end{figure}

\clearpage
\section{CONCLUSION} \label{sec:conclusion}

In this work we have systematically explored the orbital dynamics of an inertialess neutrally-buoyant deformable prolate capsule in unbounded simple shear flow using direct simulations. For a capsule with small bending stiffness, we revealed that the orbit always converges towards a unique stable equilibrium state independent of the initial orientation. As $\Ca$ increases, the stable orbit evolves from the $z$ (vorticity) axis to the shear plane. This trend holds qualitatively for all values of $\lambda$ and AR considered in this study. Four dynamical modes for the stable orbit, namely, rolling, wobbling, oscillating-swinging, and swinging, are determined at increasing $\Ca$, similar to the findings by Dupont \emph{et al.} \cite{Dupont2013}. The spontaneous curvature is shown to have a minor effect on the orbital dynamics of the capsule, although we believe that the artificial capsules in experiments are more likely to adopt a spontaneous shape the same as or close to the rest shape, instead of a flat sheet, which would lead to higher stress and strain energy in the membrane surface.

For a capsule with large bending stiffness, we found that the viscosity ratio $\lambda$ plays a significant role in the determination of the stable orbits. When $\lambda \lesssim 1$, the stable orbit is always in the shear plane. Two regimes of stable orbital motions are identified: a rigid-body-like tumbling motion at low $\Ca$, and a fluid-like swinging motion at high $\Ca$. When $\lambda > 1$, the stable motion is tumbling and rolling at low and high $\Ca$ regimes, respectively, independent of the initial orientation. During the transition, however, the capsule is found to adopt multiple stable orbital modes including tumbling, precessing and rolling, depending on the initial orientation. This multiplicity regime becomes broader as the aspect ratio of the capsule increases, while showing an opposite dependency on the viscosity ratio. We also predicted a general shear-thinning behavior and a multiplicity in the rheological properties for a dilute suspension of prolate capsules all assuming the same orbit, as a result of the multiplicity in the attractors for the capsule dynamics. 
  
\clearpage 
\begin{acknowledgments}
This work was supported by NIH Grant No.~R21MD011590-01A1.
\end{acknowledgments}

\end{document}